\documentclass[sigconf]{acmart}

\usepackage{multirow}
\usepackage{booktabs} 
\setcopyright{rightsretained}

\usepackage{graphicx}
\usepackage{caption}
\usepackage{subcaption}
\usepackage{amsmath}

\usepackage{url}
\makeatletter
\g@addto@macro{\UrlBreaks}{\UrlOrds}
\usepackage{xcolor}

\newcommand{\MediaRank}{{\em MediaRank}}
\newcommand{\NuzzelRank}{{\em NuzzelRank}}
\newcommand{\swallow}[1]{}
\newcommand{\fullversion}[1]{}

\usepackage{pifont}
\usepackage[percent]{overpic}

\usepackage{hyperref} 
\usepackage{amsthm}
\theoremstyle{definition}

\def\BibTeX{{\rm B\kern-.05em{\sc i\kern-.025em b}\kern-.08emT\kern-.1667em\lower.7ex\hbox{E}\kern-.125emX}}

\copyrightyear{2019}
\acmYear{2019}
\setcopyright{acmlicensed}
\acmConference[KDD '19]{KDD '19: 25th ACM SIGKDD Conference on Knowledge Discovery and Data Mining}{August 4 -- 8, 2019}{Anchorage, Alaska USA}

\begin{document}

%

\title{MediaRank: Computational Ranking of Online News Sources}

\author{Junting Ye}
\affiliation{%
\institution{Stony Brook University}
\city{Stony Brook}
\country{NY}}
\email{juyye@cs.stonybrook.edu}

\author{Steven Skiena}
\affiliation{%
\institution{Stony Brook University}
\city{Stony Brook}
\country{NY}}
\email{skiena@cs.stonybrook.edu}

\begin{abstract}
In the recent political climate, the topic of news quality has drawn attention both from the public and the academic communities. The growing distrust of traditional news media makes it harder to find a common base of accepted truth. In this work, we design and build \MediaRank\ (\url{www.media-rank.com}), a fully automated system to rank over 50,000 online news sources around the world. \MediaRank\ collects and analyzes one million news webpages and two million related tweets everyday. We base our algorithmic analysis on four properties journalists have established to be associated with reporting quality: peer reputation, reporting bias/breadth, bottomline financial pressure, and popularity.

Our major contributions of this paper include: \textit{(i)} Open, interpretable quality rankings for over 50,000 of the world's major news sources.   Our rankings are validated against 35 published news rankings, including French, German, Russian, and Spanish language sources. \MediaRank\ scores correlate positively with 34 of 35 of these expert rankings. \textit{(ii)} New computational methods for measuring influence and bottomline pressure. To the best of our knowledge, we are the first to study the large-scale news reporting citation graph in-depth. We also propose new ways to measure the aggressiveness of advertisements and identify social bots, establishing a connection between both types of bad behavior. \textit{(iii)} Analyzing the effect of media source bias and significance. We prove that news sources cite others despite different political views in accord with quality measures. However, in four English-speaking countries (US, UK, Canada, and Australia), the highest ranking sources all disproportionately favor left-wing parties, even when the majority of news sources exhibited conservative slants.
\end{abstract}

\maketitle
\section{Introduction}

A common base of accepted truth is perhaps the most important foundations of democracy, yet this has come under assault in our era of fake news and the widespread distrust of traditional media.
Considerable work has been 
devoted to developing NLP-based methods to detect unreliable news articles
\cite{perezrosas2018automatic,potthast2017stylometric}, as well as independent third-party fact checking services
like \textit{PolitiFact} and \textit{Snopes}, but validity checking on the article level is too brittle and slow relative to the demands of the news cycle.

We believe that the proper level to assess news quality is at the source
level, through aggregate analysis of their coverage, content, and reputation.
While the professional journalists offer accurate annotations on evaluating the quality of news sources (e.g. \textit{NewsGuard}), it is difficult and expensive for them to achieve high coverage due to the sheer amount of information generated everyday.

Towards this end, we have developed \MediaRank\ (\url{www.media-rank.com}),
a fully automated system to rank over fifty thousand online news sources around the world. We collect and analyze about one million new webpages and two million related Tweets everyday.  This longitudinal dataset represents a substantial
academic resource for analyzing news media and information flow around the world.

Ranking online news sources proves a challenging task. A straightforward approach one might use is traditional website ranking algorithms, e.g. \textit{PageRank} \cite{page1999pagerank}.
But as we will show in Section \ref{subsubsec:citation_pagerank}, this does not prove an effective approach because of ``sponsored articles'' and other uninformative hyperlinks that dominate news pages.
Instead,  multiple metrics must be considered to assess the media quality.
According to surveys of top U.S. journalists conducted by \textit{Pew Research Center}, political balance journalism, quality of coverage (e.g. depth and context) and bottomline pressure are among the key factors influencing the quality of news sources \cite{plasser2005hard}. 

\begin{table}[!t]
\small
\centering
\begin{tabular}{@{}|c|ccccc|@{}} \hline 
	\multirow{2}{*}{Rankings} 	&  Media  &  Nuzzel           & News               & Feed   & AllYou \\
	& Rank   & Rank\footnotemark & Guard\footnotemark &  Spot\footnotemark  & CanRead\footnotemark \\  \hline 
	Public & \ding{52} & \ding{56} & \ding{52} & \ding{52} & \ding{52} \\
	Multi-Topics & \ding{52} & \ding{52} & \ding{52} & \ding{52} & \ding{52} \\
	Multi-Lang & \ding{52} & \ding{52} & \ding{56} & \ding{52} & \ding{52} \\
	Multi-Nation & \ding{52} & \ding{52} & \ding{56} & \ding{52} & \ding{52} \\
	>50K Sources & \ding{52} & \ding{52} &  \ding{56} & --- & \ding{56} \\
	Interpretable & \ding{52} & --- & \ding{52} & \ding{56} & \ding{56} \\
	Algorithmic & \ding{52} & --- & \ding{56} & \ding{56} & \ding{56} \\
	\hline 
\end{tabular}
\caption{Comparisons of \MediaRank\ against other news ranking systems: \NuzzelRank, \textit{NewsGuard}, \textit{FeedSpot}, and \textit{AllYouCanRead}.  Blank entries reflect lack of reliable information concerning methodology and coverage.
	\label{tab:adv_enum}}
\vspace{-.35in}
\end{table}

\addtocounter{footnote}{-4} 

\stepcounter{footnote}\footnotetext{\scriptsize \url{https://nuzzel.com/rank}} 
\stepcounter{footnote}\footnotetext{\scriptsize \url{https://www.newsguardtech.com/}}
\stepcounter{footnote}\footnotetext{\scriptsize \url{https://www.feedspot.com/}}
\stepcounter{footnote}\footnotetext{\scriptsize \url{https://www.allyoucanread.com/}}

With this domain wisdom in mind, we propose the following four properties to assess the quality of news sources, and develop novel algorithmic methods to evaluate them:

\begin{itemize}
\item \textit{Peer Reputation}: Reliable news sources are trusted by other reliable news sources.   Reporting citations are common in online news articles. We argue that news sources receive more citations from good places have higher reputation. Therefore, we use \textit{PageRank} scores on reporting citation graph to evaluate the importance of news sources.  This metric proves to be particularly effective for large-scale news sources.
\item \textit{Reporting Bias and Breadth}: Reliable news sources strive to be politically unbiased in their search for truth.  Further, they strive to cover the full breadth of important news rather than repeated coverage of narrow domains.
We measure reporting bias by the sentiment differences towards a large universe of people associated with left- and right-wing parties.
The magnitude of sentiment bias can be accurately quantified through longitudinal analysis over a large news corpus. Breadth of reporting is estimated by the count of unique celebrities' names mentioned in their articles. 
\item \textit{Bottomline Pressure}: The business environment for news venues has become increasingly challenging, with most sources facing considerable
financial pressure to attract and monetize readers.
But bottomline pressure is regarded by journalists as the biggest concern affecting news quality \cite{plasser2005hard}. We propose two new metrics to assess integrity under financial pressure: \textit{(i)} the use of social network bots hired to boost user traffic,  and \textit{(ii)} the number and placement of ads shown on news pages to gain revenue.
\item \textit{Popularity}:  More reliable news sources are recognized as such by readers and other news sources.  Social media and content analysis links and Alexa rank scores\footnote{\url{https://www.alexa.com/siteinfo}} reflect the popularity among news readers and sources.  We demonstrate that popularity correlates strongly with peer reputation but is independent of bias.
\end{itemize}

\begin{table}[!t]
\footnotesize
\centering
\setlength{\tabcolsep}{0.4em}
\begin{tabular}{@{}|c|rrrr||c|rrrr|@{}} \hline 
	\multicolumn{5}{|c||}{News \MediaRank\ Favors} & \multicolumn{5}{c|}{News \NuzzelRank\ Favors} \\  \hline &&&&&&&&&\\[-2ex]
	News & $\Delta$ & $\widetilde{MR}$ & NR & MR & News & $\Delta$ & NR & $\widetilde{MR}$  &  MR \\  \hline 
	
	variety.com & 76 & 16 & 92 & 17 & mediamatters.org & -52 & 42 & 94 & 4290 \\
	nature.com & 62 & 14 & 76 & 15 & qz.com & -47 & 37 & 84 & 1005 \\
	sciencemag.org & 56 & 37 & 93 & 51 & gizmodo.com & -46 & 13 & 59 & 121 \\
	rollingstone.com & 46 & 45 & 91 & 65 & fastcompany.com & -44 & 26 & 70 & 211 \\
	independent.co.uk & 45 & 24 & 69 & 29 & thedailybeast.com & -44 & 25 & 69 & 204 \\
	telegraph.co.uk & 42 & 20 & 62 & 21 & entrepreneur.com & -40 & 35 & 75 & 316 \\
	apnews.com & 39 & 21 & 60 & 25 & propublica.org & -40 & 38 & 78 & 381 \\
	usatoday.com & 38 & 10 & 48 & 10 & venturebeat.com & -38 & 59 & 97 & 6521 \\
	scmp.com & 37 & 50 & 87 & 87 & zdnet.com & -36 & 30 & 66 & 177 \\
	espn.com & 37 & 8 & 45 & 8 & bostonglobe.com & -35 & 19 & 54 & 91 \\
	
	\hline 
\end{tabular}
\caption{Contrasting the top 10 news sources with biggest ranking gaps between \MediaRank\ (MR) and \NuzzelRank\ (NR). $\widetilde{MR}$ is induced \MediaRank\ value among the 97 available news from NR.}
\label{tab:mr_vs_nr}
\vspace{-.3in}
\end{table}

\MediaRank\ combines scores from the signals described above to compute a quality score for over 50,000 sources.
Table \ref{tab:adv_enum} compares our methodology \MediaRank\ to other new
ranking systems, establishing us as the only large-scale, international, algorithmic news ranking system with publicly released rankings for evaluation and
analysis.
Table \ref{tab:mr_vs_nr} compares our source rankings to \NuzzelRank,
perhaps the most comparable system, but one that releases only the relative
rankings of its top 99 sources.
Although there is general agreement (Spearman correlation 0.52) the
differences are revealing when we identify the most disparate rankings among
their sources.
We strongly prefer the sources of record in science ({\em Nature} and {\em Science}) and entertainment ({\em Variety}, {\em Rolling Stone}, and {\em ESPN}),
and professional news sources (the {\em Associated Press}, {\em Independent},
and {\em Telegraph}), while \NuzzelRank\ favors blog-oriented sources like
{\em VentureBeat}, {\em QZ}, and {\em Media Matters}.

\swallow{
comparing the top 10 news sources from \MediaRank\ and \textit{NuzzelRank}. We noticed that \NuzzelRank\ gives \textit{BBC} 19th ranking, while \textit{The Atlantic} and \textit{Forbes} get 3rd and 4th place. We argue that \textit{BBC} should be ranked higher because it has 5th highest reputation score (\textit{BBC}'s reputation score 0.433, popularity 0.9997, breadth 0.25. The larger these metrics are, the better a source is. The other scores are the opposite). \textit{The Atlantic} and \textit{Forbes} get much lower reputation, popularity and coverage quality scores comparing to other top news outlets (\textit{The Atlantic}'s reputation 0.097, popularity 0.9923, breadth 0.169; \textit{Forbes}'s reputation 0.143, popularity 0.9983, breadth 0.211; ).
}

\begin{figure}[!t]
	\centering
	\begin{tabular}{cc}
		\hspace{-.2in}
		\includegraphics[width=0.5\linewidth]{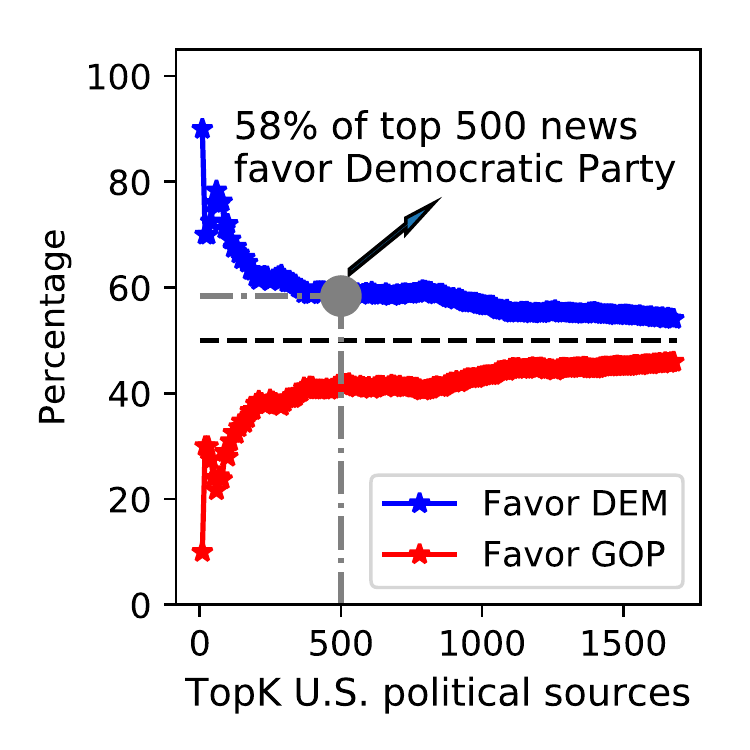} &
		\hspace{-.1in}
		\includegraphics[width=0.5\linewidth]{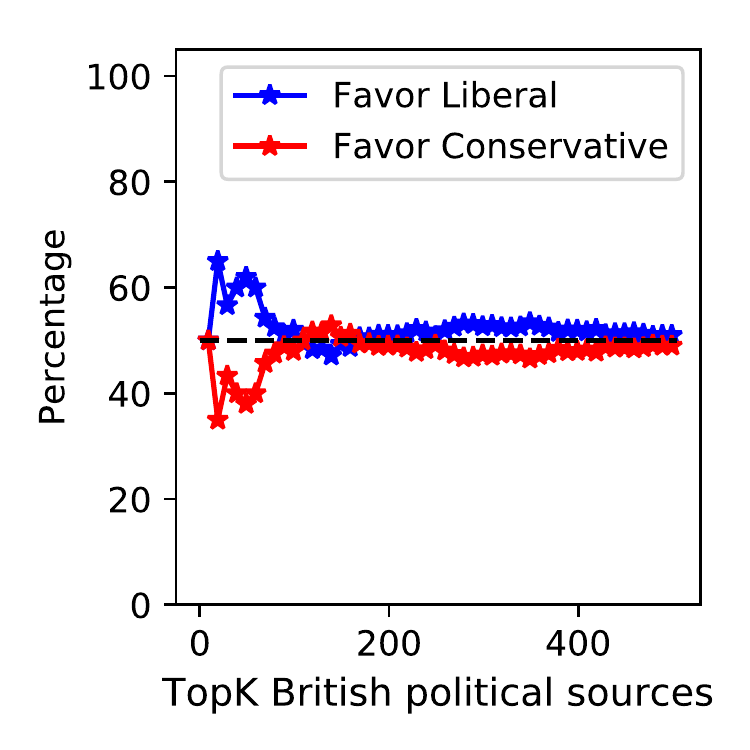}\\
		\hspace{-.2in}
		\includegraphics[width=0.5\linewidth]{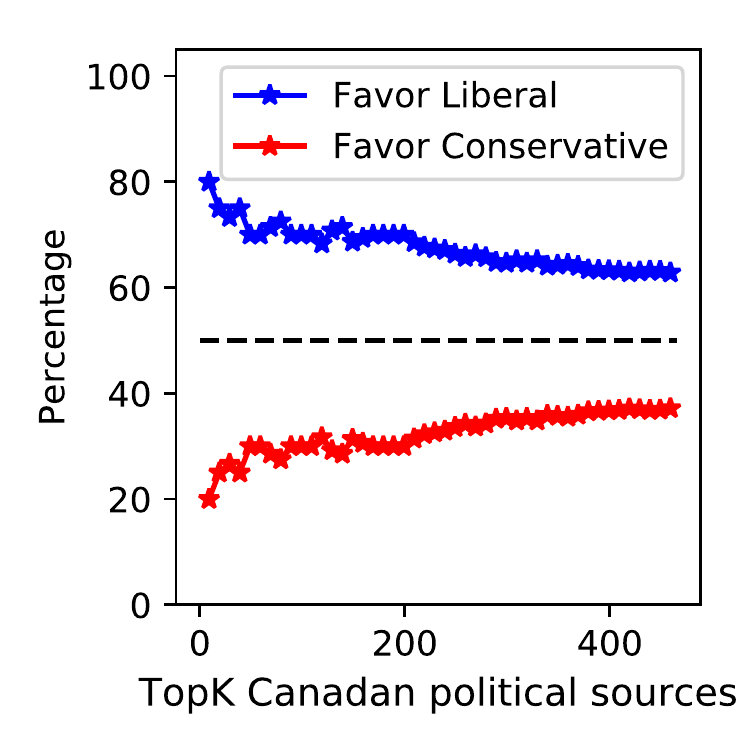} &
		\hspace{-.1in}
		\includegraphics[width=0.5\linewidth]{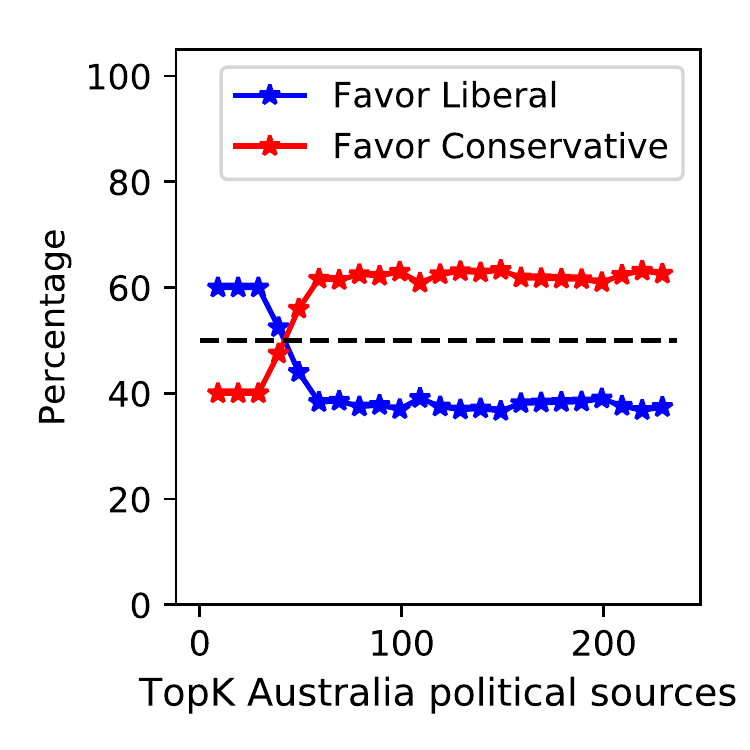}
	\end{tabular}
	\vspace{-.1in}
	\caption{{Bias of national news towards liberal and conservative parties in major English-speaking countries (U.S., U.K., Canada and Australia). The best news sources tend to favor liberal parties in all these countries.}
		\label{fig:localNewsBias}}
	\vspace{-.15in}
\end{figure}

The major contributions in this paper are:

\begin{itemize}
\item {\em Open, interpretable quality rankings for the world's major news sources} --
We provide detailed computational analysis for over 50,000 news sources from around the world.
We evaluate our rankings against 35 published news rankings, including French,
German, Russian, and Spanish language sources.  \MediaRank\ scores correlate positively with 34 of 35 of these expert rankings, achieving a mean Spearman coefficient of 0.58.  We concur with 24 of these expert rankings at above a 0.05-significance level, with a mean coefficient 0.69.
Each source ranking score can be interpreted by six intuitive metrics regarding reputation, popularity, quality of coverage and bottomline pressure. 
We will make this analysis fully available to the research community
and general public at \url{www.media-rank.com}.

\item {\em New computational methods for measuring influence and bottomline pressure/social bots} --
To the best of our knowledge, we are the first to study the large-scale news reporting citation graph in-depth.
We are also the first to study computational ways to measure the bottomline pressure among news sources. Observing online news make most of their revenue from user traffic and online advertisement, we propose methods to detect social bots that promotes website traffic and to track the volume and aggressiveness of advertisements on news webpages.
These metrics present interesting views into the business of the media world,
and new tools for analyzing other websites and social media properties.

\item {\em Media bias and significance} -- We have performed
extensive experiments using our signal metrics to quantify properties of media sources, with interesting results. In particular, we prove that news sources cite others despite different political views (Figure \ref{fig:citationgraphembedding}) in accord with quality measures.  We also were surprised to learn that neutral sources were {\em not} those most highly ranked by other metrics.
Indeed, in four English-speaking countries (US, UK, Canada, and Australia), the
highest ranking sources all disproportionately favor left-wing parties, even when the majority of news sources exhibited conservative slants
(Figure \ref{fig:localNewsBias}).

\end{itemize}


\section{Related Work}
\label{sec:relatedwork}

The problem of news source ranking has been attracting growing attention from academic and industrial researchers.
Corso et. al. studied the problem of simultaneously ranking news sources and its stream of news articles \cite{del2005ranking}.
They proposed a graph formulation where nodes are news sources and articles.
The edges reflect relations between sources and articles, and content similarity between articles.
A time-aware label propagation algorithm is proposed to assign weights to nodes in this graph. 
Mao and Chen suggested a similar approach to simultaneously rank news sources, topics and articles, assuming that trust-worthy news sources publish high-quality articles concerning important news topics \cite{mao2010method}. 
Hu et. al. analyzed the visual layout information of news homepages to exploit the mutually reinforcing relationship between news articles and news sources \cite{hu2006discovering}.
These methods are dependent on computationally expensive models over articles, like label propagation. Therefore they are limited to small news corpora, and not appropriate for datasets with hundreds of millions of articles like ours.

\NuzzelRank\ is a news recommendation system which also generates rankings of news sources.
They claim their scores are computed by combining the reading behavior of their users, the engagement and authority of news sources and signals from news reliability initiatives such as the \textit{Trust Project}\footnote{\scriptsize \url{https://thetrustproject.org}} and \textit{NewsGuard}\footnote{\scriptsize \url{https://www.newsguardtech.com}}.
We identified their top 99 ranked news sources (all that they made available to the public as of Oct. 23, 2018) for
comparison with \MediaRank. 

Online misinformation is now drawing increased attention from the research community \cite{perezrosas2018automatic, potthast2017stylometric, ruchansky2017csi, shu2017fake, varol2017online}.
Zhang et.al. define credibility indicators in news articles for manual annotation, including eight content (e.g. title representativeness, quotes from outside experts, etc.) and eight context indicators (e.g. originality, representative citations, etc.) \cite{zhang2018structured}.
Linguistic models achieve limited performance in detecting fake news, especially the ones aim to deceive readers \cite{perezrosas2018automatic}.
A hybrid model combining news text, received responses and the source users promoting them is proposed by \cite{ruchansky2017csi}.
Online misinformation spreads quickly on social media platforms, due to the convenience of message sharing \cite{shao2018spread}.
Algorithms designed to take down social bots who publish or share misinformation or other content automatically include \cite{varol2017online,cresci2017paradigm,cresci2015fame,lee2011seven}.

Substantial efforts have been made to analyze and rank individual news articles by information retrieval community \cite{ter2018faithfully, kong2012ranking}.
Kiritoshi and Ma rank news articles by estimating the relatedness, diversity, polarity and detailedness of its named entities \cite{kiritoshi2014named}.
Tatar et. al. uses user comments to predict the popularity of news articles \cite{tatar2014popularity}.
 Godbole et. al. propose efficient algorithms for large-scale sentiment analysis of online news and social media \cite{godbole2007large}.
Kulkarni et. al. design a multi-view attention model to classify the political ideology of news articles \cite{vivek2018multiview}. 

\section{\MediaRank\ Overview}
\label{sec:mediarank}

The lack of valid ground-truth labels makes news ranking a challenging task. In this work, we design effective and interpretable component signals from different perspectives regarding news quality. This makes it easy to explain why one source is better than the other. Considering the sheer amount of news data everyday,  each signal metric we use has been designed so be scalable for large-scale data analysis.

\MediaRank\ is a large system, with 1 master server and 100 dedicated slave servers processing the world's news.
It is organized in following four major components:

\begin{enumerate}
\item \textit{News source discovery}:
two strategies are employed to identify new sources:
i). new URLs appear on Google News, and  ii) new URLs appear in Tweets returned by \textit{Twitter API} when searching with keyword ``news''.
Between Sep. 24, 2017 and Oct. 30, 2018, 50,834 unique news sources are discovered in this way, with 87\% of our tracked sources being from \textit{Google News}.
The remaining 13\% sources identified from Twitter prove less well-known,
 but sometimes go viral in social media.

\item \textit{Collecting news webpages and related tweets}:
We use \textit{Newspaper3k}\footnote{\scriptsize \url{https://github.com/codelucas/newspaper}} to collect and parse news webpages from discovered domains.
We also extract URLs from collected tweets to see whether it is tracked in \MediaRank. If yes, we further query its user profile data from Twitter and keep them for analysis.
On average, \MediaRank\ collects about one million raw HTMLs and two million news related tweets each day. A cluster of 20 machines performs data collection and cleaning.

\item \textit{Analysis and News Ranking}: Multiple signals have been shown to be correlated with the quality of news sources, including reputation among peers, the degree of political bias, and popularity among readers.
We devote a cluster of 80 machines to computation-intensive analysis, including named entity recognition, sentiment analysis, social bots classification, and duplicate article detection.

\item \textit{Visualization and API}: Our goal is to make \MediaRank\ an important data source to support external research efforts in journalism and the social sciences as well as computer science.
We are designing APIs to provide online service, notifying Web users whether the news they consume are from low quality sources.
\end{enumerate}

\begin{figure}[!t]
	\centering
	\begin{tabular}{c}
		\hspace{-.1in}
		\includegraphics[width=0.95\linewidth]{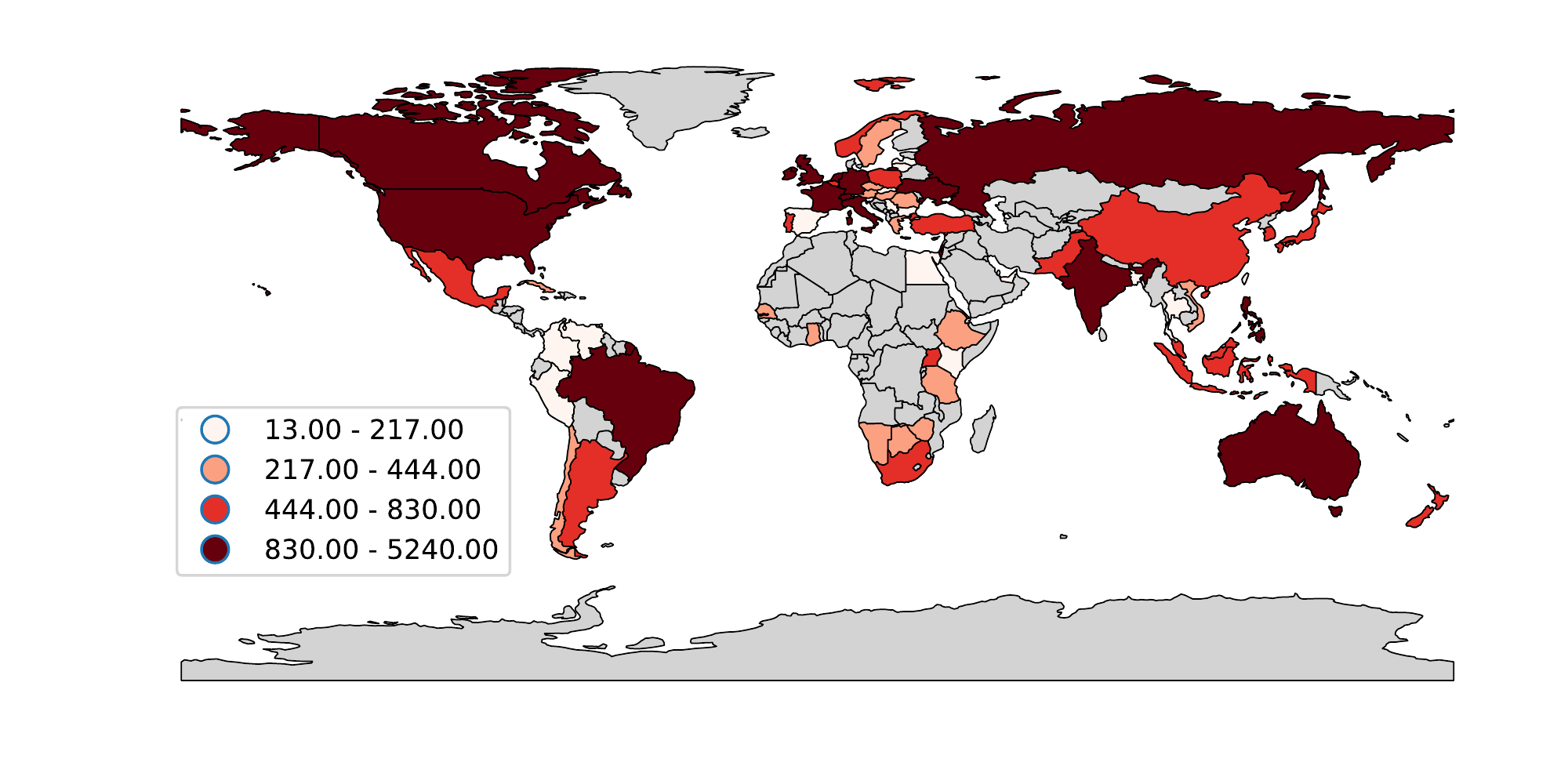}
	\end{tabular}
	\vspace{-.15in}
	\caption{{\MediaRank\ tracks 50,696 news sources from 68 countries. Colors represent the number of sources per country. 5,240 sources are from United States. Countries with zero tracked news sources are marked in grey.}
		\label{fig:locDistri}}
	\vspace{-.1in}
\end{figure}

Figure \ref{fig:locDistri} shows the national distribution of tracked sources, using meta data from \textit{Google News}.
We observe that most sources are from western countries, with limited data from Africa and Middle East. Fully five thousand sources are from United States.
Italy, Russia, Canada and U.K. are next four countries in terms of
source frequency.
Only 36\% of our sources publish in English.
Multi-language sources like \textit{BBC} are labeled as per which language is used in the most articles. Sources with multiple topics, like the \textit{New York Times}, are labeled as ``General''. 

\section{News Citations}
\label{sec:citation}

Just as academic papers cite other papers,
online news articles often acknowledge their peers' work as information sources.
We argue that such citations can generally be viewed as endorsements among journalism peers.
To the best of our knowledge, we are the first to generate large-scale news citation graphs for in-depth analysis and news ranking.

In this section, we analyze citation behavioral patterns of news sources (Table \ref{tab:citationPatters}).
We also define the news citation graph, where the nodes are news sources and directed edges represent citations between source pairs.

\subsection{Dataset}
\label{subsubsec:Data}

\fullversion{
\begin{figure}[!t]
\centering
\begin{tabular}{cc}
	\begin{subfigure}[b]{0.42\columnwidth}
		\hspace{-.2in}
		\includegraphics[width=1\textwidth]{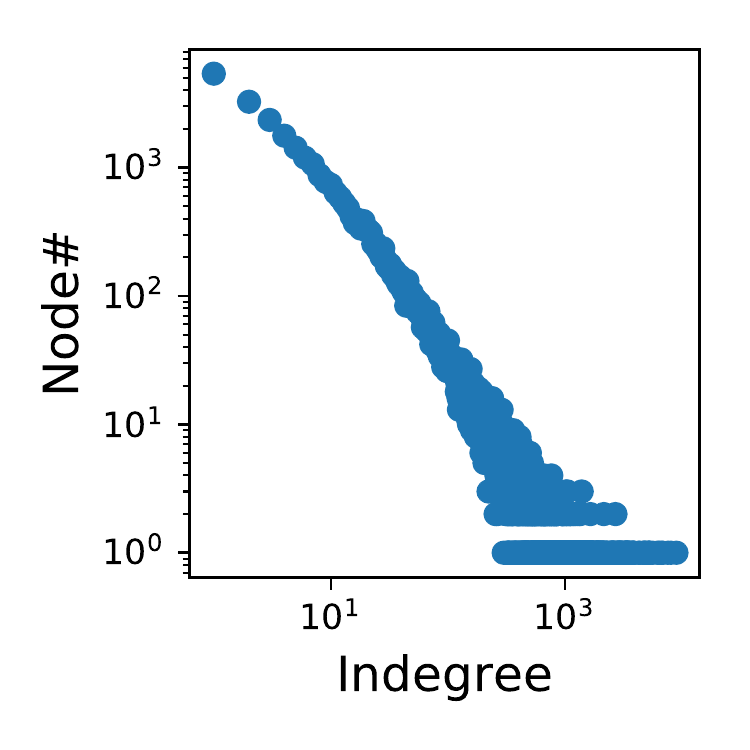}
		\vspace{-.1in}
	\end{subfigure}
	\begin{subfigure}[b]{0.42\columnwidth}
		\hspace{-.2in}
		\includegraphics[width=1\textwidth]{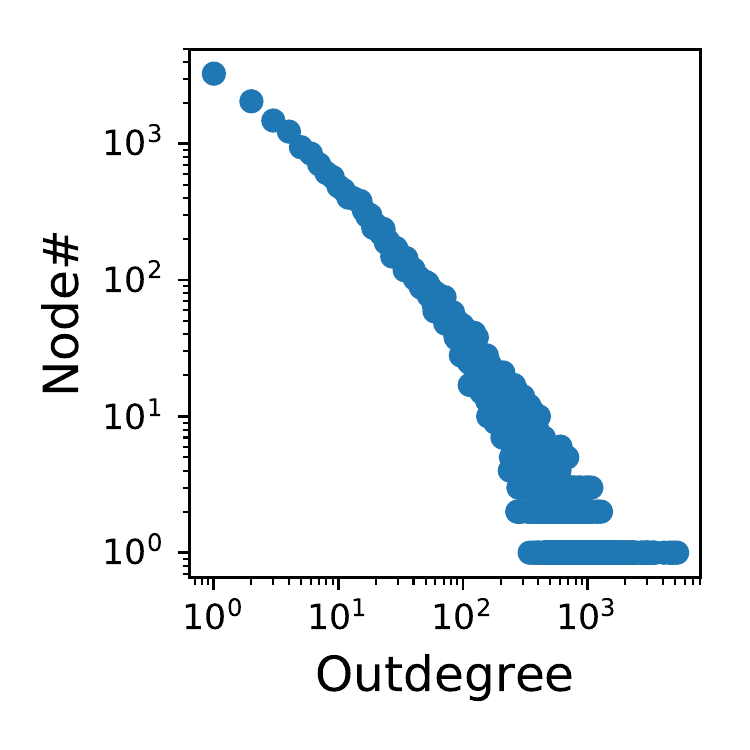}
		\vspace{-.1in}
	\end{subfigure}\\
	\begin{subfigure}[b]{0.42\columnwidth}
		\hspace{-.2in}
		\includegraphics[width=1\textwidth]{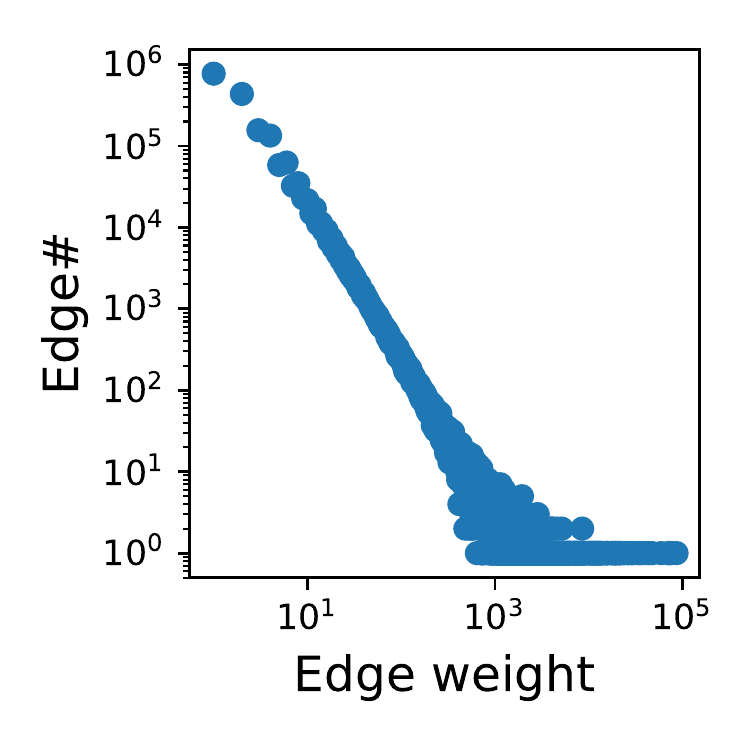}
		\vspace{-.15in}
	\end{subfigure}
	\begin{subfigure}[b]{0.42\columnwidth}
		\hspace{-.2in}
		\includegraphics[width=1\textwidth]{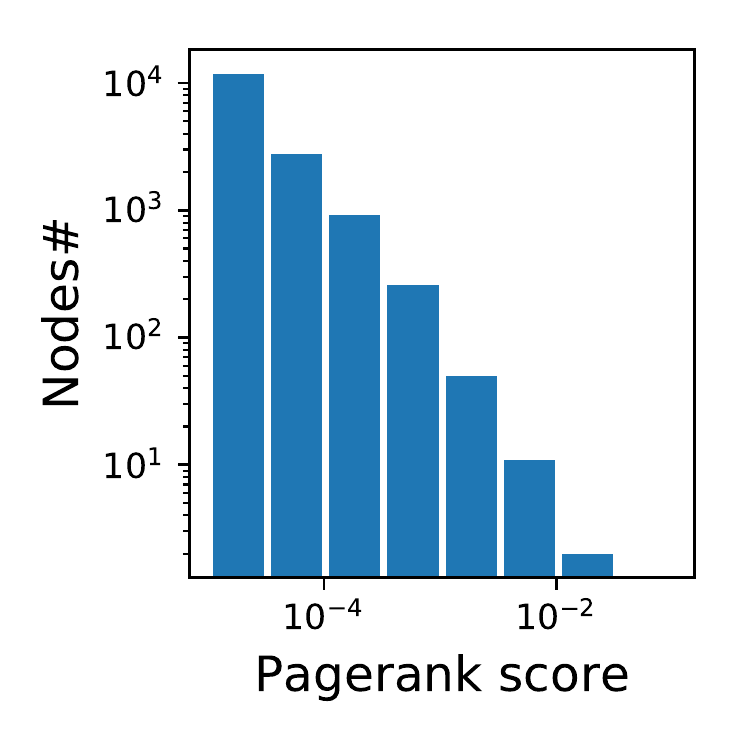}
		\vspace{-.15in}
	\end{subfigure}
\end{tabular}
\caption{{Indegree, outdegree, edge weight and pagerank score distributions of news citation graph. All distributions show linearity in log-log scale, indicating popular news received exponentially more citations from their peers.}
	\label{fig:citationDegree}}
\vspace{-.15in}
\end{figure}}

We analyze 23,371,264 articles collected between Sep. 24, 2017 and Feb. 16, 2018.
Each article contains at least one citation inside it, for a total of
64,976,942 citations.
Of these, 42,734,224 (66\%) are self citations to given news source,
while 22,242,718  (34\%) cite different news sources.

\fullversion{
As shown in Figure \ref{fig:citationDegree}, the distributions of indegree, outdegree, edge weights and pagerank show linear curvature, resembling power-law distribution. Most nodes (sources) have few neighbors (citing other sources or being cited by others), while small amount of nodes attract many links from other. For example, news with largest in-degrees, ``nytimes.com'', ``theguardian.com'' and ``bbc.com'' are cited by 8928, 8020 and 7540 other news sources, respectively. 

It is also interesting to see from edge weight distribution (left bottom subgraph in Figure \ref{fig:citationDegree}) that, there are 122 edges have more weights larger than 10,000 (i.e. more than 10,000 citations from one source to another). It turns out that 45 of them are local community websites of ``sbnation.com'' having large amount of links directing to ``sbnation.com'', which makes sense, e.g. ``bleedinggreennation.com'' (a \textit{Philadelphia Eagles} community), ``cincyjungle.com'' ( a \textit{Cincinnati Bengals} community), ``ninersnation.com'' (a \textit{San Francisco 49ers} community), etc. }

The news citation graph is a directed graph, denoted as $G_c = <V, E, W>$
where news sources are the nodes, $V$. $e_{ij} \in E$ is a directed edge from node $v_i$ to $v_j$ ($v_i, v_j \in V$).
The total number of citations from $v_i$ to $v_j$ defines
$w_{ij} \in W$ the weight of edge $e_{ij}$, 
%
Our weighted source citation graph contains 50696 nodes and
1,947,189 edges after removing self-loop edges.


\begin{table}[!t]
	\small
	\centering
	\setlength{\tabcolsep}{0.4em}
	\begin{tabular}{@{}|l|rrrr|@{}} \hline 
	\MediaRank\ &	\multirow{2}{*}{doc/day}        &      	\multirow{2}{*}{$C_s^{out}$/doc}           &       \multirow{2}{*}{$C_o^{out}$/doc}           &      \multirow{2}{*}{$C_o^{in}$/doc}            \\
	 tier &  &  &  &   \\  \hline
$[$1,500) & 47.0  &  2.8  &  1.0  &  201.0 \\
$[$500, 2K) & 17.3  &  2.0  &  0.9  &  60.3 \\
$[$2K, 5K) & 9.0  &  1.7  &  0.9  &  18.4 \\
$[$5K, 10K) & 5.1  &  1.6  &  0.9  &  8.8 \\
$[$10K, 20K) & 3.3  &  1.5  &  0.8  &  1.8 \\
$[$20K, 50K) & 2.3  &  1.3  &  0.8  &  0.2 \\
		\hline 
	\end{tabular}
	\caption{Higher ranking news sources \textit{(i)} publish more articles each day, \textit{(ii)} have more citations to both articles of their own and other sources, and \textit{(iii)} receive more citations from others.
Sources are grouped into six tiers based on their \MediaRank\ values.
$C_s^{out}$: count of self-citations, $C_o^{out}$: citations to other sources, $C_o^{in}$: citations from others.  }
	\label{tab:citationPatters}
		\vspace{-.15in}
\end{table}

\subsection{Citation Ranking}
\label{subsubsec:citation_pagerank}

\textit{PageRank} was famously defined as an algorithm to rank websites \cite{page1999pagerank}. The key idea is that every webpage propagates their weight to their neighbors.
When a page has many links from large-weight webpages, the weight of this page increases.
Similarly, we argue that citations between news sources should be interpreted as endorsements among journalists.
When a news source is disproportionately cited by its peers, it indicates a higher journalistic reputation. 

We compare \textit{PageRank} results on both citation graph and URL graph (where sources are connected by all URLs, instead of just inside articles.).
By comparing the top 10 news from both rankings,
we observed that certain sources ranked disturbingly higher in the URL graph than in citation graph.
For example, ``digg.com'' stands 6th on URL ranking, while only 71st in citation ranking.
``bna.com'' is placed 8th on URL ranking vs. 1157th on citation ranking.
The primary reason for such anomalies is that outside article links are often ads or ``sponsored'' articles, which prove much less informative than reporting citations. 

We use \textit{PageRank} values from the citation graph to quantify peer reputation, normalized to be in the range $[0,1]$.
The greater the reputation score is, the better the source is presumed to be.

\subsection{Citation News Embeddings}
\label{subsec:ciationNewsEmbed}

\begin{figure}[!t]
	\centering
	\begin{tabular}{c}
		\hspace{-.05in}
		\includegraphics[width=1\linewidth]{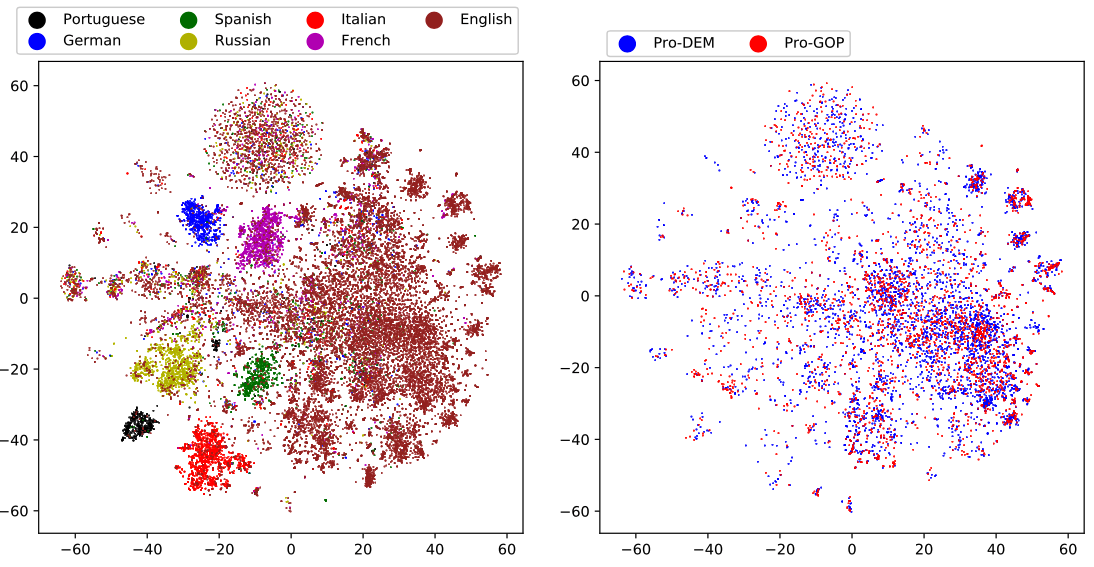}
	\end{tabular}
	\vspace{-.1in}
	\caption{{2D projection of news embeddings learned from the citation graph. (Left): the colors are labeled based on news sources' languages. Strong clusters are formed by all major languages. (Right): the colors are labeled based on political news sources' sentiment towards U.S. parties. No large clusters are observed, indicating that news sources cite each other despite different political views.}
		\label{fig:citationgraphembedding}}
	\vspace{-.15in}
\end{figure}

Graph embeddings are low-dimension vector representations for nodes so that similar nodes have similar representations \cite{perozzi2014deepwalk}.
We are interested in how news sources align in embedding space, and what their nearest neighbors look like.
We used \textit{Node2Vec} \cite{grover2016node2vec} to learn news sources embeddings on news citation graph, and projected these embeddings
into two dimensions for visualization purpose using \textit{t-SNE} \cite{van2014accelerating}.
For visualization purposes, we used meta-data from \textit{Google News} to label sources by topics and annotated sources by political bias as
explained in Section \ref{sec:newsbias}.

Figure \ref{fig:citationgraphembedding} (left) shows the news sources distribution of top seven languages tracked in \MediaRank.
All languages form strong clusters.
English proves widely used in many countries,
so we see multiple smaller national sub-clusters.
For the right figure, it is interesting that no large clusters are found among political news sources. This indicates that sources with different political views do cite each other, contradicting the ``echo chamber'' effect associated with social media platforms \cite{flaxman2016filter}.

\section{News Bias}
\label{sec:newsbias}

News bias is a critical metric reflecting the quality of news sources.
According to \textit{Pew Research Center} survey of 38 countries, a median of 75\% per nation say it is never acceptable for a news organization to favor one political party over others \cite{amy2018publics}. 


To facilitate large-scale text analysis, we employ efficient and effective algorithms to extract named entities \cite{finkel2005incorporating} and compute sentence-level sentiment \cite{gilbert2014vader}. The political bias of a new source is computed by aggregating sentiments towards party members. 


\subsection{Datasets}
%

We analyzed news articles collected from Sep. 24th, 2017 to Dec. 31, 2018,
with 427,464 distinct celebrities are mentioned at least once.
There are 77,596,029 articles containing at least one celebrity's name, totaling 614,440,328 mentions.
These celebrities' English names are mapped to entities extracted from \textit{DBpedia Data Set 3.1}\footnote{\scriptsize \url{https://wiki.dbpedia.org/data-set-31}}. 
We identified each celebrities' political party label from
DBpedia, with 2,908 unique parties are associated with 58,131 celebrities, of which
 12,784 are U.S. Republicans and 11,774 Democrats.
We have enriched this with Trump's cabinet (past) members\footnote{\scriptsize \url{https://en.wikipedia.org/wiki/Cabinet_of_Donald_Trump}}, 115th\footnote{\scriptsize \url{https://en.wikipedia.org/wiki/115th_United_States_Congress}} and 116th\footnote{\scriptsize \url{https://en.wikipedia.org/wiki/116th_United_States_Congress}} class of congress members for analysis.

We identified two external resources to prove a ground-truth for news bias evaluation:
\begin{itemize}
\item
\textit{AllSides}\footnote{\scriptsize \url{https://www.allsides.com/media-bias/media-bias-ratings}}: 222 raw news sources with their political bias.
Each source is labeled with one of following five political views by news editors: \textit{left}, \textit{left-center}, \textit{center}, \textit{right-center} and \textit{right}, these labels are also voted by Web users.
After filtering out those not tracked by \MediaRank, 117 news sources remained. We also observed that there is often an inconsistency between the opinions of news editors and Web users.
We removed the inconsistent sources and those labeled as ``center'', leaving 
71 news sources for evaluation.

\item
\textit{MediaBiasFactCheck} (\textit{MBFC})\footnote{\scriptsize \url{https://mediabiasfactcheck.com}}: contains 1040 news sources labeled as ``Left Bias'', ``Left-center Bias'', ``Right-center Bias'' and ``Right Bias''.
Of these, 653 are tracked in \MediaRank.
We combined ``Left Bias'' and ``Left-center Bias'' in news as ``Left'' and
``Right-center Bias'' and ``Right Bias'' as ``Right''.

\end{itemize}

\subsection{Sentiment Aggregation}
\label{subsec:sentimentaggregation}
We now explain the details of how the sentiment of news entities and sources are computed.
We consider three ways to aggregate news sentiment:
\begin{itemize}
\item
{\em Article-level bias Vote (AV)}: each article has one vote towards an entity: positive, negative or neutral. The group sentiment is aggregated by counting votes from articles containing a party member.
\item 
{\em Article-level bias Distribution (AD)}: similar to AV, but aggregating entity sentiment distributions instead of votes.
\item
{\em Sentence-level bias Distribution (SD)}: similar to SD, but assigning weights proportional to entity mentions instead of articles.
\end{itemize}

Formally, we assume news source $s_i = <d_1, d_2, ..., d_n>$ consists of a sequence of articles.
Each article, $d_j = <g_1, g_2,.., g_m>$, consists of a sequence of sentences.
Let
$E_k = <e_1, e_2, .., e_u>$ denote the list of entities occurring
 in sentence $k$.
Let $O(g_k)$ denote the sentiment probability distribution of sentence $g_k$. The distribution has three classes, positive, neutral and negative sentiments. For example, $O(g_k) = [0.1, 0.9, 0.0]$, where the entries are positive, neutral and negative sentiment scores, respectively.
For each entity, its party affiliation $P(e_l), e_l \in E_k$, can be one of the 2,908 parties or none. The average sentiment distribution of party $p_u$ from article $d_j$ is defined:

\begin{equation}
\label{equ:sent_ent}
O(d_j, p_u) =\frac{1}{N_d} \sum_{g_k \in d_j} \sum_{e_l \in E_k} O(g_k) * I(P(e_l) = p_u)
\end{equation}
where $N_d$ is the normalization term that makes $O(d_j, p_u)$ a probability distribution. $I(\cdot)$ is an indicator function, whose value is one if the condition is satisfied, otherwise 0. $O(g_k)$ is viewed as vector when under adding or multiplying operations.
An article's sentiment towards a political party is the average sentiment of its sentences.
$\overline{O}(d_j, p_u)$ denotes the vote of article $d_j$ on party $p_u$. This one-hot vector denote whether it is a positive, neutral or negative sentiment vote. For example, $\overline{O}(d_j, p_u) = [0, 1, 0]$ is a neutral vote if the positive of $O(d_j, p_u)$ equals the negative. It takes $[1,0,0]$ if the positive of $O(d_j, p_u)$ is larger than the negative sentiment, otherwise $[0,0,1]$.

\begin{table}[!t]
	\small
	\centering
	\begin{tabular}{@{}|c|cc|cc|@{}} \hline 
		Method & EntitySet & Entity\# & AllSides & MBFC  \\  \hline
		Random & -- & -- & 0.489 & 0.502 \\ \hline
		Article Vote & Cabinet & 27 & 0.507 & 0.509 \\
		Article Distri. & Cabinet & 27 & 0.493 & 0.540 \\
		Sentence Distri. & Cabinet & 27 & 0.541 & 0.526 \\
		Article Vote & Congress & 564 & 0.701 & 0.540 \\
		Article Distri.  & Congress & 564 & 0.656 & 0.530 \\
		Sentence Distri. & Congress & 564 & 0.666 & 0.557 \\
		Article Vote & All & 18773 & 0.761 & 0.643 \\
		Article Distri. & All & 18773 & 0.746 & 0.649 \\
		\textbf{Sentence Distri.} & \textbf{All} & \textbf{18773} & \textbf{0.764} & \textbf{0.683} \\
		\hline 
	\end{tabular}
	\caption{Accuracies of news source bias prediction. MBFC: MediaBiasFactCheck.com.}
	\label{tab:biasEstimation}
	\vspace{-.15in}
\end{table}

We aggregate the article-level vote as:
\begin{equation}
\label{equ:sent_AV}
O_{av}(s_i, p_u) = \frac{1}{N_{av}} \sum_{d_j \in s_i} \overline{O}(d_j, p_u)
\end{equation}
where $N_{av}$ is the normalization term that makes $O_{av}(s_i, p_u)$ a probability distribution.
The article-level aggregate distribution $O_{ad}(n_i, p_u)$ is defined similarly using $O(d_j, p_u)$.
\swallow{

is defined as Equation \ref{equ:sent_AD}.

\begin{equation}
\label{equ:sent_AD}
O_{ad}(s_i, p_u) = \frac{1}{N_n} \sum_{d_j \in n_i} O(d_j, p_u)
\end{equation}

}
The sentence-level aggregate distribution is computed:
\begin{equation}
\label{equ:sent_SD}
O_{sd}(s_i, p_u) =\frac{1}{N_{sd}} \sum_{d_j \in s_i} \sum_{g_k \in d_j} \sum_{e_l \in E_k} O(g_k) * I(P(e_l) = p_u)
\end{equation}

Finally,the sentiment score a news source for a political party is computed:
\begin{equation}
\label{equ:bias_score}
B(s_i, p_u) = \frac{O^{pos}(s_i, p_u) - O^{neg}(s_i, p_u)}{O^{pos}(s_i, p_u) + O^{neg}(s_i, p_u)} 
\end{equation}
where $O^{pos}(s_i, p_u)$ and $O^{neg}(s_i, p_u)$ are the positive and negative values of sentiment distribution $O(s_i, p_u)$. $B(s_i, p_u)$ is in the range $[-1, 1]$. The absolute gap between sentiment scores of left- or right-wing parties is used to quantify source bias.


\subsection{News Bias Evaluation}
\label{subsec:newsbiaseval}

To evaluate our methods for political bias detection, we used source bias labels from two organizations, \textit{AllSides} and \textit{MediaBiasFactCheck} (i.e. \textit{MBFC}),  as ground-truth data. Table \ref{tab:biasEstimation} shows how our various sentiment methods perform using different groups of party-associated entities.
Accuracy increases when using larger sets of party-associated entities for all aggregation methods.
SD aggregation slightly outperforms other methods.

\begin{table}[!t]
	\small
	\centering
	\setlength{\tabcolsep}{0.65em}
	\begin{tabular}{@{}|l|cr|cr||c|@{}} \hline 
		\multirow{2}{*}{News} & \multicolumn{2}{c|}{Democratic} & \multicolumn{2}{c||}{Republican} & \multirow{2}{*}{MBFC Label}  \\ \cline{2-5}
		&   Bias & \#(K) & Bias & \#(K) &   \\ \hline
		latimes.com & \textcolor{blue}{+0.06} & 93 & +0.00 & 264 & left-center\\
		businessinsider.com & \textcolor{blue}{+0.07} & 87 & +0.00 & 256 & left-center\\
		theconversation.com & \textcolor{blue}{+0.13} & 5 & +0.05 & 17 & center\\
		fortune.com & \textcolor{blue}{+0.10} & 15 & +0.05 & 44 & right-center\\
		smh.com.au & \textcolor{blue}{+0.10} & 17 & +0.03 & 51 & left-center\\
		usnews.com & \textcolor{blue}{+0.10} & 37 & +0.05 & 119 & left-center\\
		vice.com & \textcolor{blue}{+0.04} & 9 & -0.01 & 31 & left-center\\
		indiatimes.com & \textcolor{blue}{+0.20} & 3 & +0.08 & 13 & left-center\\
		qz.com & \textcolor{blue}{+0.12} & 8 & +0.06 & 22 & left-center\\
		miamiherald.com & \textcolor{blue}{+0.08} & 10 & +0.01 & 21 & left-center\\
		\hline
		sky.com & -0.14 & 7 & \textcolor{red}{-0.06} & 21 & left-center*\\
		breitbart.com & +0.03 & 140 & \textcolor{red}{+0.09} & 380 & extreme-right\\
		nationalreview.com & +0.02 & 44 & \textcolor{red}{+0.08} & 98 & right\\
		dailycaller.com & +0.02 & 58 & \textcolor{red}{+0.08} & 129 & right\\
		torontosun.com & -0.11 & 8 & \textcolor{red}{+0.00} & 18 & right \\
		eveningtimes.co.uk & -0.05 & 4 & \textcolor{red}{+0.04} & 10 & --\\
		clarionledger.com & +0.06 & 13 & \textcolor{red}{+0.15} & 44 & --\\
		abc7.com & +0.02 & 5 & \textcolor{red}{+0.09} & 15 & --\\
		dailyecho.co.uk & -0.04 & 2 & \textcolor{red}{+0.01} & 8 & --\\
		nationalinterest.org & +0.05 & 5 & \textcolor{red}{+0.12} & 27 & right-center\\
		\hline 
	\end{tabular}
	\caption{Successfully discriminating the ten most significant  left- and right-wing news sources by sentiment.
*Note that \textit{Sky News} is owned by \textit{21st Century Fox},
and considered a conservative source by \textit{Wikipedia}.}
	\label{tab:newsbias_left}
		\vspace{-.15in}
\end{table}

\fullversion{
\begin{table}[!t]
	\small
	\centering
	\begin{tabular}{@{}|l|cr|cr||c|@{}} \hline 
		\multirow{2}{*}{News} & \multicolumn{2}{c|}{Democratic} & \multicolumn{2}{c||}{Republican} & \multirow{2}{*}{MBFC Label}  \\ \cline{2-5}
		&   Bias & \#(K) & Bias & \#(K) &   \\ \hline
		sky.com & -0.14 & 7 & -0.06 & 21 & left-center*\\
		breitbart.com & +0.03 & 140 & +0.09 & 380 & extreme-right\\
		nationalreview.com & +0.02 & 44 & +0.08 & 98 & right\\
		dailycaller.com & +0.02 & 58 & +0.08 & 129 & right\\
		torontosun.com & -0.11 & 8 & +0.00 & 18 & right \\
		eveningtimes.co.uk & -0.05 & 4 & +0.04 & 10 & -- \\
		clarionledger.com & +0.06 & 13 & +0.15 & 44 & -- \\
		abc7.com & +0.02 & 5 & +0.09 & 15 & -- \\
		dailyecho.co.uk & -0.04 & 2 & +0.01 & 8 & --\\
		nationalinterest.org & +0.05 & 5 & +0.12 & 27 & right-center\\
		\hline 
	\end{tabular}
	\caption{Sentiment of top 20 right-wing news sources. Sky News is seen as a right-wing news outlet due to its ownership by 21st Century Fox, according to Wikipedia.}
	\label{tab:newsbias_right}
\end{table}}

Table \ref{tab:newsbias_left} presents the most significant left and right-leaning news sources, where the gap between democratic and republican bias $> 0.05$.
There is excellent agreement with \textit{MBFC} bias labels.
The outlier is \textit{Sky News}
labeled as left-center by \textit{MBFC} but owned by \textit{21st Century Fox},
and considered a conservative source by \textit{Wikipedia}\footnote{\scriptsize \url{https://en.wikipedia.org/wiki/Sky_News}}.

\definecolor{c0}{RGB}{29, 145, 192}
\definecolor{c1}{RGB}{33, 149, 192}
\definecolor{c2}{RGB}{36, 152, 193}
\definecolor{c3}{RGB}{40, 156, 193}
\definecolor{c4}{RGB}{43, 160, 194}
\definecolor{c5}{RGB}{47, 163, 194}
\definecolor{c6}{RGB}{51, 167, 194}
\definecolor{c7}{RGB}{54, 171, 195}
\definecolor{c8}{RGB}{58, 175, 195}
\definecolor{c9}{RGB}{61, 178, 196}
\definecolor{c10}{RGB}{65, 182, 196}
\definecolor{c11}{RGB}{71, 184, 195}
\definecolor{c12}{RGB}{77, 187, 194}
\definecolor{c13}{RGB}{84, 189, 193}
\definecolor{c14}{RGB}{90, 191, 192}
\definecolor{c15}{RGB}{96, 194, 192}
\definecolor{c16}{RGB}{102, 196, 191}
\definecolor{c17}{RGB}{108, 198, 190}
\definecolor{c18}{RGB}{115, 200, 189}
\definecolor{c19}{RGB}{121, 203, 188}
\definecolor{c20}{RGB}{127, 205, 187}
\definecolor{c21}{RGB}{134, 208, 186}
\definecolor{c22}{RGB}{141, 211, 186}
\definecolor{c23}{RGB}{149, 213, 185}
\definecolor{c24}{RGB}{156, 216, 184}
\definecolor{c25}{RGB}{163, 219, 183}
\definecolor{c26}{RGB}{170, 222, 183}
\definecolor{c27}{RGB}{177, 225, 182}
\definecolor{c28}{RGB}{185, 227, 181}
\definecolor{c29}{RGB}{192, 230, 181}
\definecolor{c30}{RGB}{199, 233, 180}
\definecolor{c31}{RGB}{203, 234, 180}
\definecolor{c32}{RGB}{207, 236, 179}
\definecolor{c33}{RGB}{210, 238, 179}
\definecolor{c34}{RGB}{214, 239, 179}
\definecolor{c35}{RGB}{218, 240, 178}
\definecolor{c36}{RGB}{222, 242, 178}
\definecolor{c37}{RGB}{226, 243, 178}
\definecolor{c38}{RGB}{229, 245, 178}
\definecolor{c39}{RGB}{233, 246, 177}
\definecolor{c40}{RGB}{237, 248, 177}
\definecolor{c41}{RGB}{239, 249, 181}
\definecolor{c42}{RGB}{241, 249, 185}
\definecolor{c43}{RGB}{242, 250, 189}
\definecolor{c44}{RGB}{244, 251, 193}
\definecolor{c45}{RGB}{246, 252, 197}
\definecolor{c46}{RGB}{248, 252, 201}
\definecolor{c47}{RGB}{250, 253, 205}
\definecolor{c48}{RGB}{251, 254, 209}
\definecolor{c49}{RGB}{253, 254, 213}
\definecolor{c50}{RGB}{255, 255, 217}
\definecolor{c51}{RGB}{255, 255, 216}
\definecolor{c52}{RGB}{255, 255, 214}
\definecolor{c53}{RGB}{255, 255, 213}
\definecolor{c54}{RGB}{255, 255, 212}
\definecolor{c55}{RGB}{255, 255, 210}
\definecolor{c56}{RGB}{255, 255, 209}
\definecolor{c57}{RGB}{255, 255, 208}
\definecolor{c58}{RGB}{255, 255, 207}
\definecolor{c59}{RGB}{255, 255, 205}
\definecolor{c60}{RGB}{255, 255, 204}
\definecolor{c61}{RGB}{255, 253, 200}
\definecolor{c62}{RGB}{255, 251, 195}
\definecolor{c63}{RGB}{255, 250, 191}
\definecolor{c64}{RGB}{255, 248, 186}
\definecolor{c65}{RGB}{255, 246, 182}
\definecolor{c66}{RGB}{255, 244, 178}
\definecolor{c67}{RGB}{255, 242, 173}
\definecolor{c68}{RGB}{255, 241, 169}
\definecolor{c69}{RGB}{255, 239, 164}
\definecolor{c70}{RGB}{255, 237, 160}
\definecolor{c71}{RGB}{255, 235, 156}
\definecolor{c72}{RGB}{255, 233, 152}
\definecolor{c73}{RGB}{255, 231, 147}
\definecolor{c74}{RGB}{255, 229, 143}
\definecolor{c75}{RGB}{254, 227, 139}
\definecolor{c76}{RGB}{254, 225, 135}
\definecolor{c77}{RGB}{254, 223, 131}
\definecolor{c78}{RGB}{254, 221, 126}
\definecolor{c79}{RGB}{254, 219, 122}
\definecolor{c80}{RGB}{254, 217, 118}
\definecolor{c81}{RGB}{254, 213, 114}
\definecolor{c82}{RGB}{254, 209, 110}
\definecolor{c83}{RGB}{254, 205, 105}
\definecolor{c84}{RGB}{254, 201, 101}
\definecolor{c85}{RGB}{254, 197, 97}
\definecolor{c86}{RGB}{254, 194, 93}
\definecolor{c87}{RGB}{254, 190, 89}
\definecolor{c88}{RGB}{254, 186, 84}
\definecolor{c89}{RGB}{254, 182, 80}
\definecolor{c90}{RGB}{254, 178, 76}
\definecolor{c91}{RGB}{254, 174, 74}
\definecolor{c92}{RGB}{254, 171, 73}
\definecolor{c93}{RGB}{254, 167, 71}
\definecolor{c94}{RGB}{254, 163, 70}
\definecolor{c95}{RGB}{254, 160, 68}
\definecolor{c96}{RGB}{253, 156, 66}
\definecolor{c97}{RGB}{253, 152, 65}
\definecolor{c98}{RGB}{253, 148, 63}
\definecolor{c99}{RGB}{253, 145, 62}
\definecolor{c100}{RGB}{253, 141, 60}

\definecolor{e1}{RGB}{228,26,28}
\definecolor{e2}{RGB}{55,126,184}
\definecolor{e3}{RGB}{77,175,74}
\definecolor{e4}{RGB}{152,78,163}
\definecolor{e5}{RGB}{255,127,0}
\definecolor{e6}{RGB}{166,86,40}
\definecolor{blue}{RGB}{0,0,255}
\definecolor{green}{RGB}{0,128,0}

\definecolor{grey}{RGB}{128,128,128}
\definecolor{k}{RGB}{0,0,0}
\definecolor{g}{RGB}{77,175,74}
\definecolor{r}{RGB}{228,26,28}

\fullversion{
Table \ref{tab:celebritySentiment} shows top 20 celebrities with most mentions from democratic and republican parties. The celebrities with positive sentiment bias is colored with green, while negative with red. Donald Trump and Bret Kavanagh get neutral sentiment and marked with grey. We can see most celebrities get positive bias, while the negatives are more likely to be involved in scandals or resignations. It is also interesting to see non-political celebrities get extremely positive sentiment, e.g. Operah Winfrey (+0.45) and Warren Buffett (+0.47). The celebrity with most negative sentiment is Harvey Weinstein. He was accused of sexual harassment by several actress, which leads to \#MeToo movement across the U.S.}

\section{Social Bot Score}
\label{sec:newsinsocial}
Social media has become the primary vehicle for news consumption:
62\% of U.S. adults received news on social media in 2016\footnote{\scriptsize \url{http://www.journalism.org/2016/05/26/news-use-across-social-media-platforms-2016/}}.
Social media outperforms television as the primary news source for younger generation (18 to 24 year old)\footnote{\scriptsize \url{www.bbc.com/news/uk-36528256}}.
Unfortunately, social media has also become the major outlet for distributing fake news \cite{ruchansky2017csi}, because the ``echo chamber'' effect makes fake news seem more trust-worthy \cite{schmidt2017anatomy}.
Social bots are social media accounts controlled by computer programs. They are often used to promote public figures by following them, or to boost business by sharing related posts. It was reported that up to 15 percent of Twitter accounts are in fact bots rather than people \cite{bot2017icwsm}. In this section, we will elaborate on how we train a social bot classifier and further compute the social bot score of news sources.

\subsection{Dataset}
\label{subsubsec:DataCollection}
Twitter is one of the most popular social media platforms, and provides
an API\footnote{\scriptsize \url{https://developer.twitter.com/en/docs.html}} enabling us to identify the 
user ID, tweet content, related URL, and post timestamp for millions of tweets.
We used the keyword ``news'' in API queries to identify news-oriented tweets,
and extracted all news-oriented URLs from these tweets.
Between Sep. 29, 2017 and Oct. 30, 2019 (397 days), we collected 715,050,598 tweets with URLs, of which 347,164,578 (48.6\%) contain URLs from tracked news sources.
These Tweets are posted by 32,275,806 users, whose profiles are also collected for social bot identification.


We identified two datasets of social bot labels for training and evaluation:
\begin{itemize}
	\item \textit{Botometer}: this dataset is the combination of four public social bots datasets from the research community\footnote{\scriptsize Download: \url{https://botometer.iuni.iu.edu/bot-repository/datasets.html}} \cite{varol2017online,cresci2017paradigm,cresci2015fame,lee2011seven}. Bot labels are collected using ``honeypot'' (i.e. followers of accounts that post random words), or by followers bought from companies.
This dataset contains 46,459 total accounts, split between 24,267 social bots and 22192 regular users.
	\item 
	\textit{Removed Accounts}:
Twitter strives to remove social bot accounts\footnote{\scriptsize \url{https://www.theverge.com/2018/3/11/17107192/twitter-tweetdecking-spam-suspended-accounts-mass-retweeting}}. We identified deleted accounts (enriched in bots) by retrieving the same user profiles twice (on Oct. 1st, 2017 and Mar. 21, 2018).
Among the original user set of 1,105,536 accounts, fully 45,654 (4.1\%) were not available after six months. 
\end{itemize}

\subsection{Social Bots Detection}
%
We model bots detection as a supervised classification problem, using 12 features extracted from user profiles.
Although follower and followee relations have proven useful in previous studies,
this was not feasible on \MediaRank\ scale due
to Twitter API rate limits.
The features we use are defined in Table \ref{tab:SpammerFeatures}.

\begin{table}[!t]
	\small
	\setlength{\tabcolsep}{0.4em}
	\centering
	\begin{tabular}{@{}|l|c|p{1.8in}|@{}} \hline
		\# & Feature & Note\\  \hline
		1 & $c_{er}$ & Count of followers \\
		2 & $c_{ee}$ & Count of followees \\
		3 & $ r = c_{ee} / c_{er}$ & Ratio of followee count over followers \\
		4 & $s = \log(\max(c_{er}, c_{ee}))$ & Log of follower or followee count\\
		5 & $r * s$ & Ratio times the log of follower/followee \\
		6 & $v$& Whether the user is verified\\
		7 & $c_f$ & Favourites count \\
		8 & $c_l$ & Listed count \\
		9 & $c_d$ & The length of profile description\\
		10 & Geo& Whether geo is enabled \\
		11 & Location &  Whether location is specified \\
		12 & Time zone& Whether time zone is specified\\
		13 & Default profile & Whether default profile background is changed\\
		14 & Default profile image & Whether default profile background image is changed \\
		\hline 
	\end{tabular}
	\caption{Features form social bots classification model using users' profile data.}
	\label{tab:SpammerFeatures}
	\vspace{-.25in}
\end{table}

The distribution of twitter account labels is highly imbalanced (only 4.1\% as removed).
We sampled 45,654 non-removed accounts as negatives for training.
Both datasets were split 70\% for training, 10\% for parameter tuning and 20\% for testing. 
As shown in Table \ref{tab:performances}, XGBoost consistently outperforms SVM classifier with RBF kernel and logistic regression model with ridge regularization on both datasets.

\begin{table}[!t]
	\small
	\centering
	\begin{tabular}{@{}|l|ccc|ccc|@{}} \hline
		\multirow{2}{*}{Model} & \multicolumn{3}{c|}{\textit{Botometer}} &  \multicolumn{3}{c|}{\textit{RemovedAccounts}} \\  \cline{2-7}
		&Pre. & Rec. & F1 & Pre. & Rec. & F1 \\ \hline 
		LR & 0.81 & \textbf{0.85} & 0.83 & 0.65 & \textbf{0.69} & 0.67 \\
		SVM & 0.84 & \textbf{0.85} & 0.84 & 0.75 & 0.61 & 0.67 \\
		\textbf{XGBoost} & \textbf{0.88} & 0.84 & \textbf{0.86} & \textbf{0.79} & 0.60 & \textbf{0.68} \\
		\hline 
	\end{tabular}
	\caption{Performance comparisons of logistic regression (LR), SVM and XGboost on two different social bot datasets.}
	\label{tab:performances}
		\vspace{-.25in}
\end{table}

\subsection{News Bot Scores}

Therefore, we employed XGBoost models trained on two datasets to all 32 million users to get their social bot scores.
Bot scores of news sources are computed by aggregating the scores for all related Twitter accounts.
Sources with high bot scores
likely that it hires bots to increase their visibility.

To be precise, let $b_u$ be the bot score of Twitter user $u$ and $T_i = <t_{i1}, t_{i2}, ..., t_{in}>$ denote the sequence of tweets with URLs directing to news source $s_i$.
Let $U(t)$ denote the user of tweet $t$.
Therefore, the bot score $B(s_i)$ of source $s_i$ is defined:
\begin{equation}
\label{equ:newsbotscore}
B(s_i) = \frac{1}{|T_i|} \sum_{t \in T_i} b_{U(t)}
\end{equation}
We combine the two models
from \textit{Botometer} and \textit{RemovedAccounts} by using the larger of the respective scores.

\section{Other Signals}
\label{sec:othersignals}
\subsection{Popularity}
\label{subsec:alexaRank}
\fullversion{
Alexa Rank is generated by an American web traffic analysis company, Alexa Internet, Inc. Its toolbar collects data on Internet browsing behavior and transmits them to the Alexa website, where they are stored and analyzed. This is the basis for Alexa Rank. 

}
\textit{Alexa Rank} is used to estimate news sources' popularity among news readers. We collected ranking values of all sources on Sep. 23rd, 2018,
using their API to collected data for the past 30 days.
The average ranking values of 30 days are computed to measure its popularity.



Alexa ranks range from 1 to 1,000,000, which we divide 20 equal-range tiers.
The top tier features sources with Alexa ranks between 1 to 50,000, including
6,932 (14\%) of the 50K news sources tracked by \MediaRank.

\fullversion{
\begin{figure}[!t]
	\centering
	\begin{tabular}{c}
		\includegraphics[width=0.75\linewidth]{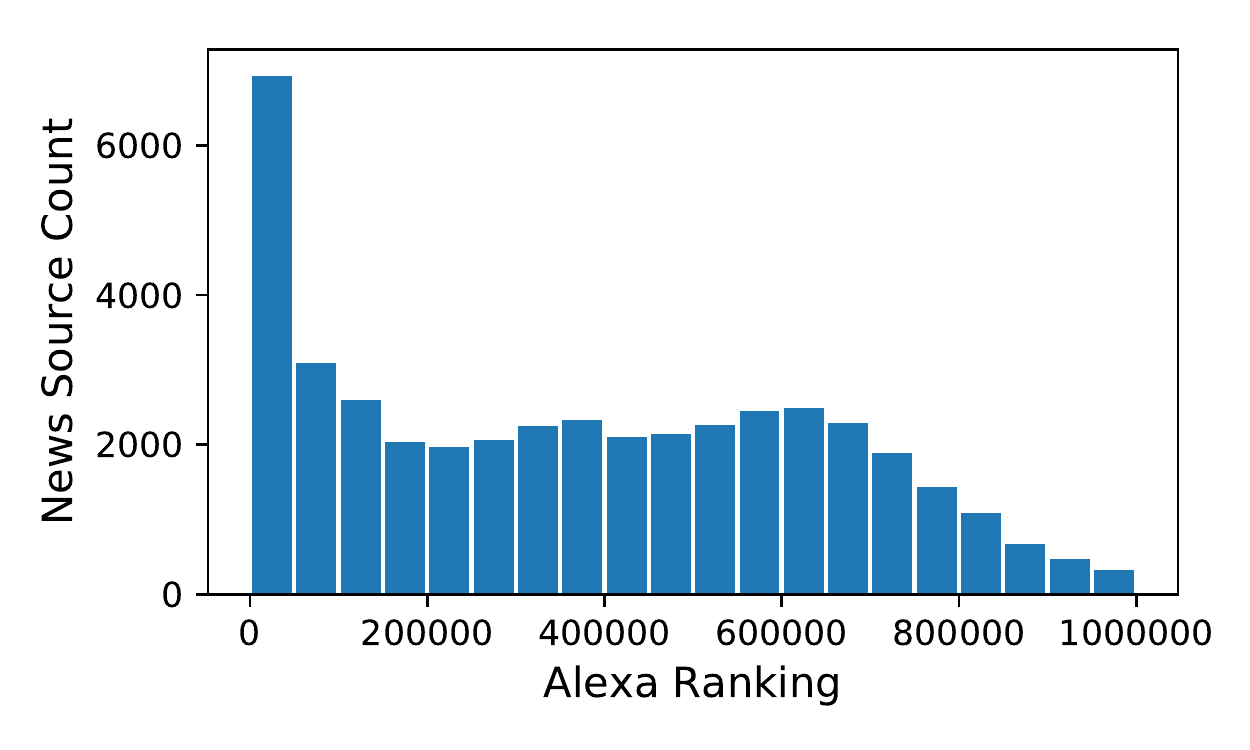}
	\end{tabular}
	\vspace{-.2in}
	\caption{{Alexa rank distribution.}
		\label{fig:alexaRanking}}
\end{figure}}

\subsection{Advertisement Aggressiveness}
\label{subsec:onlineads}

\begin{figure}[!t]
	\centering
	\begin{tabular}{c}
		\includegraphics[width=0.8\linewidth]{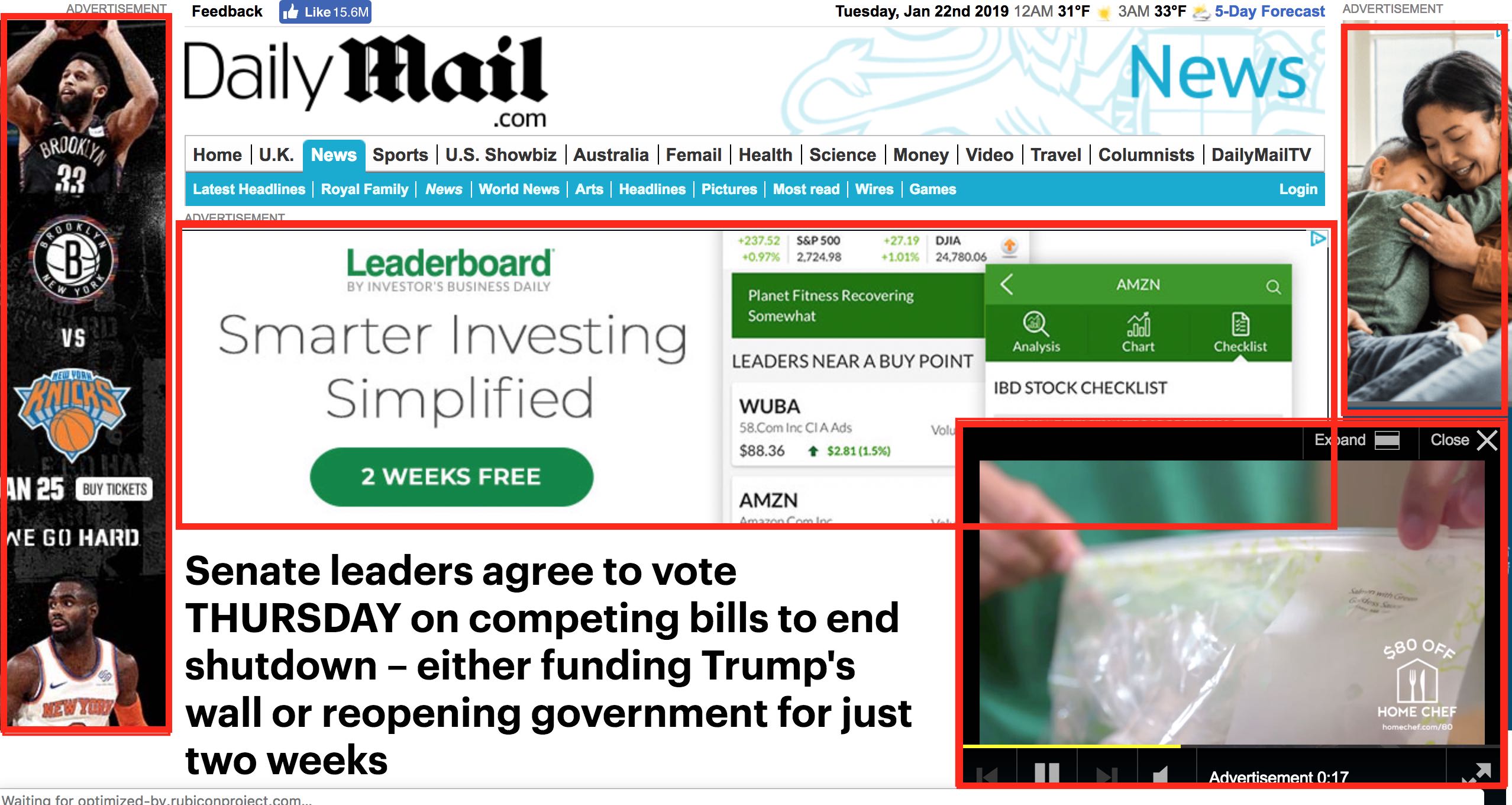}
	\end{tabular}
\vspace{-.1in}
	\caption{{The \textit{Daily Mail} is a notoriously aggressive advertiser,
here with 20 digital advertisements overwhelming the news title.}
		\label{fig:ads_page}}
	\vspace{-.2in}
\end{figure}

Online advertising is the major revenue stream for many news sources.
Media properties under great bottom-line pressure may increase the presence of ads on their pages, reducing user experience to gain more reader clicks/impressions to survive. 

\fullversion{The straightforward approach is using \textit{AdBlocker Plus}\footnote{\scriptsize \url{https://adblockplus.org/}}, a widely used \textit{Chrome} extension to block ads. We tried to make use of their list of rules to detect ads URLs, but it turned out to get substantial amount of true negatives (e.g. ads URLs not rendered on webpage).}
To collect news advertising data,
we used \textit{Selenium}\footnote{\scriptsize \url{https://www.seleniumhq.org/}} to discover rendered iFrames from \textit{Google Ads} platform in HTMLs. 
This was effective in terms of precisions, but less so in recall.

As an example, Figure \ref{fig:ads_page} shows the first screen of a webpage from the \textit{Daily Mail}, a popular British news source.
The four observable ads here are distracting, making it hard to notice the news titles on the bottom of the page.
We encountered the \textit{Daily Mail} articles with as many as twenty ads per page, making it an example of advertising aggressiveness.

\subsection{Reporting Breadth}
\label{subsec:NameEntity}

The breadth of coverage is an important indicator of news quality, reflecting the scope, relevance, depth insight, clarity, and accuracy of reporting \cite{plasser2005hard}.
We use the number of unique entities to measure the breadth of news reporting.
Good news sources strive to cover the full breadth of important news,
rather than narrow domains with limited and repeated entity occurrence.

\fullversion{
\begin{figure}[!t]
	\centering
	\begin{tabular}{c}
		\includegraphics[width=0.7\linewidth]{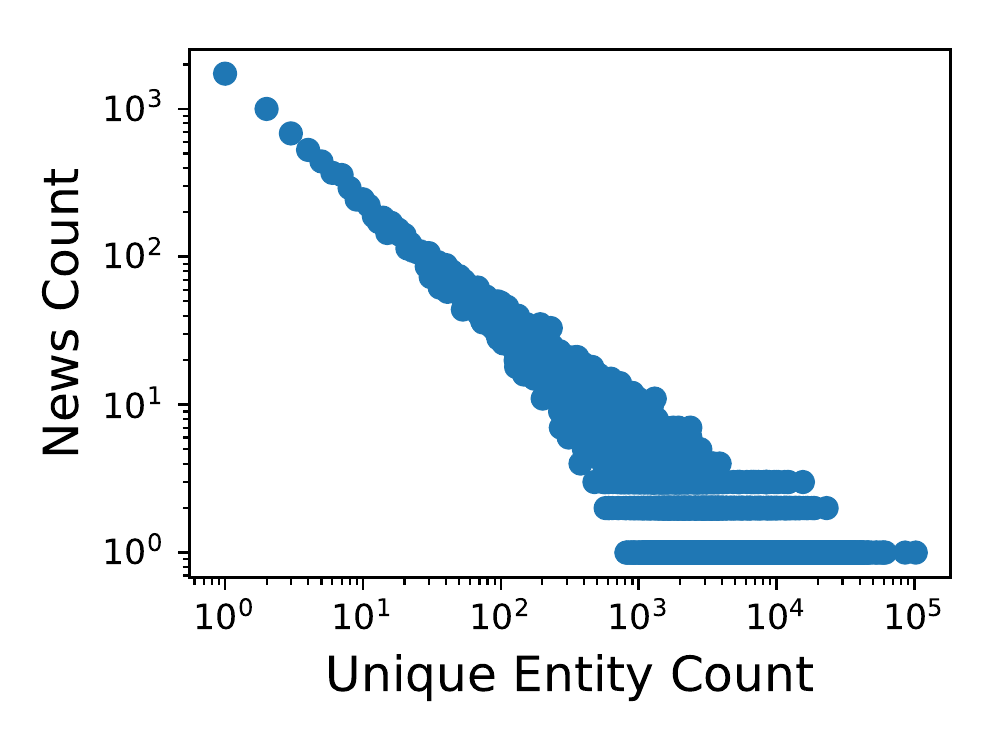}
	\end{tabular}
	\caption{{Distribution of unique entities mentioned in articles.}
		\label{fig:news_entity}}
\end{figure}}

	\section{Consensus Source Ranking}
	\label{sec:RankingAlgorithm}
	
	\subsection{Methodology}
	
	\fullversion{
	\begin{table}[!t]
		\small
		\centering
		\begin{tabular}{@{}|l|c|c|p{1.8in}|@{}} \hline
			\# & Signal & Type & Definition \\  \hline
			1 & $f_p$ & continuous & page rank scores on citation graph  \\
			2 & $f_a$ &  continuous & average of 30-day Alexa rank scores \\
			3 & $f_b$ & continuous & political bias score \\
			4 & $f_e$ & continuous & count of unique entities \\
			5 & $f_s^g$ & binary & average bot scores of related Twitter users, trained on ``ground truth'' data \\
			6 & $f_s^d$ & binary & average bot scores of related Twitter users, trained on deleted Twitter user profiles \\
			7 & $f_o$ & binary & excessive online advertisement \\
			\hline 
		\end{tabular}
		\caption{Signals Used for \MediaRank\ ranking}
		\label{tab:mediaranksingals}
			\vspace{-.25in}
	\end{table}}

	In our ranking model, each news source is represented by a vector of signal scores: reputation, popularity, reporting breadth, political bias, bot score and advertising aggressiveness, denoted
by $f_r, f_p, f_e, f_b, f_s$ and $f_a$ respectively.
For the four continuous signals $F = [f_r; f_p; f_e; f_b;]$ (each normalized in the range of [0, 1]), the source ranking score is defined:
	\begin{equation}
	R(s_i) = W^T \cdot F_i \cdot C_p^{I(f_s = 1) + I(f_a = 1)}
	\label{equ:mediarank}
	\end{equation}
where $W = [w_r; w_p; w_e; w_b;]$ is the weight vector for these
signals and $W^T$ is the transpose of $W$.
$C_p$ is the penalizing factor to discount the weights of sources employing social bots and displaying excessive ads, measured as binary (0 or 1) features using 95 percentile values as thresholds.
$I(\cdot)$ is an indicator function whose value is 1 iff the condition is satisfied.
Empirically we set the $W = [1.65; -0.35; 0.05; -0.10;]$ and $C_p = 0.95$, reflecting the monotonicity of each feature.
	
	\subsection{Evaluation}
	
	\begin{figure}[!t]
		\centering
		\begin{tabular}{c}
			\includegraphics[width=.95\linewidth]{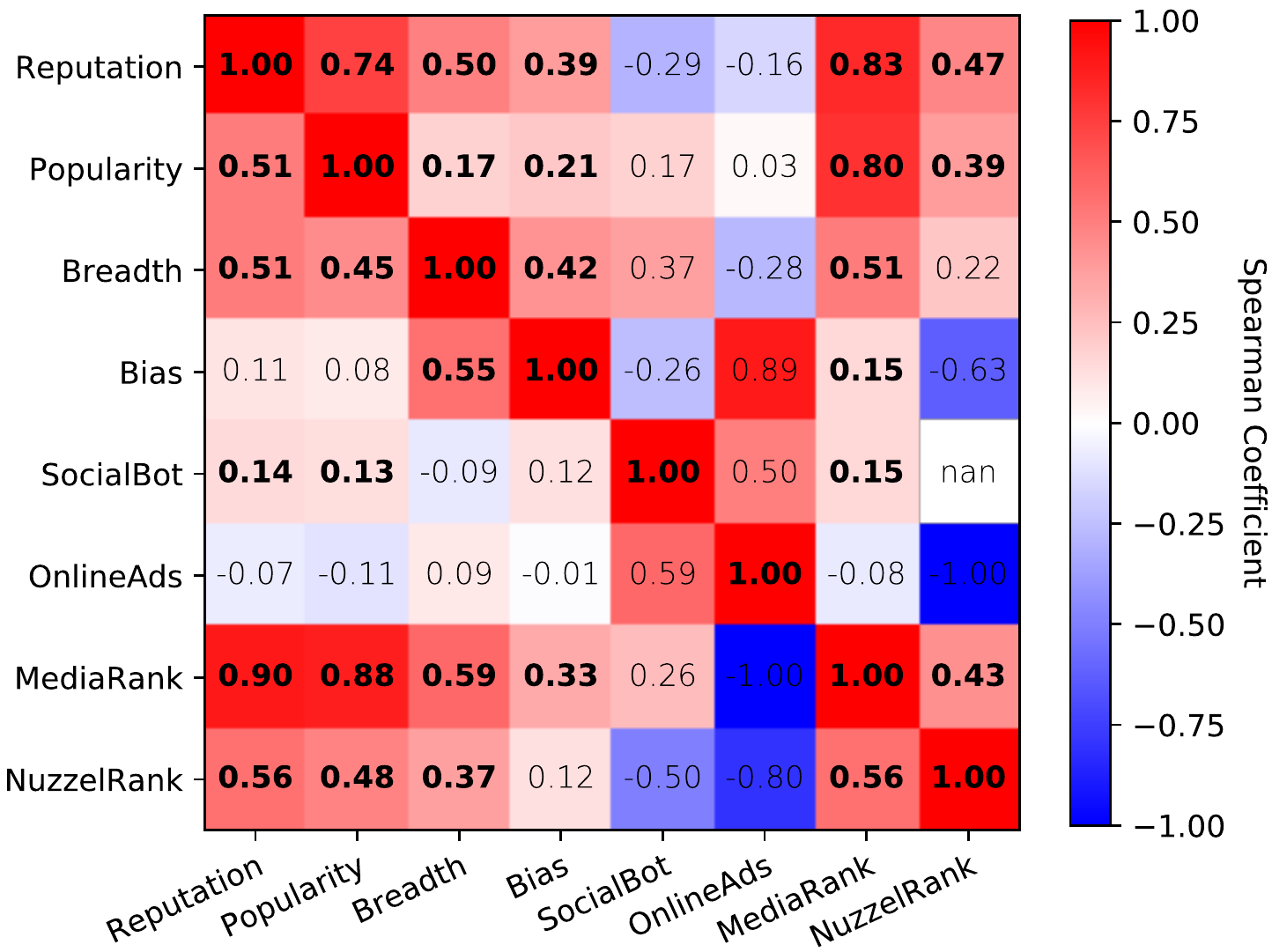}
		\end{tabular}
	\vspace{-.1in}
		\caption{{Spearman rank coefficients between pairs of signals, \MediaRank\ and \textit{NuzzelRank} on common sources. All coefficients with $> 0.05$ significance are in bold. }
			\label{fig:ranking_coef}}
		\vspace{-.15in}
	\end{figure}

Figure \ref{fig:ranking_coef} presents the Spearman rank correlation between signal pairs and two source rankings (\MediaRank\ and \textit{NuzzelRank}) on common sources.
Because vastly more low quality news sources than high quality outlets,
we use stratified sampling to compute correlations.
These samples are drawn from six news tiers as sorted by \MediaRank\ scores,
with boundaries of rank 100, 400, 1600, 6400 and 25600.
We have sampled 100 news sources from each tier.
We compare the \MediaRank\ rankings of the 600 sampled news to their \textit{NuzzelRank} rankings. When comparing \textit{NuzzelRank} to \MediaRank, the sampled news are different, thus the coefficient matrix is not symmetric.
For bot and ads scores, we use the gaps to thresholds as ranking values (news with zeros are ignored). Reputation, popularity and breadth scores highly correlate with each. Coefficients of bot, ads and \textit{NuzzelRank} prove less significant due to smaller number of associated news sources.

	In addition, we compare \MediaRank\ scores with
	35 expert news sources rankings (including French, German, Italian, Russian, and Spanish language sources).
	We also propose a ranking quality metric to quantify how good the selection of news are when comparing to \MediaRank.
	Let
		\begin{equation}
			\label{equ:rankingquality} 
			Q(S) = \sum_{s_i \in S} \frac{1}{rank_{s_i}} \ , rank_{s_i} \in [1, m]
		\end{equation}
		\begin{equation}
		\label{equ:rankingquality_norm} 
		Q_n(S) = \frac{Q(S) - Q(S_{min})}{Q(S_{max}) - Q(S_{min})} 
		\end{equation}
	where $rank_{s_i}$ is the \MediaRank\ value of news source $s_i$ among $m$ sources tracked in \MediaRank.
	$Q(S_{min})$ is the smallest value of $Q(\cdot)$ because news in $S_{min}$ stand at the bottom of \MediaRank. Similarly, $Q(S_{max})$ gets the largest value when $S_{max}$ is at the top. Therefore,
	$Q_n(S)$ is normalized in range $[0, 1]$ as a final news ranking score.
	The high quality scores observed in Table \ref{tab:comparepublicranking} demonstrates that we agree with the experts that these sources are important, not just their relative rankings as measured by Spearman correlation.

\begin{table}[!t]
	\footnotesize
	\centering
	\setlength{\tabcolsep}{0.3em}
	\begin{tabular}{@{}|c|cc|c|rrr|@{}} \hline
		\multirow{2}{*}{External Rankings}  &  \multicolumn{2}{c|}{News Group}  & \multirow{2}{*}{$T_n$/$N$} & \multicolumn{3}{c|}{Compared to \MediaRank} \\ \cline{2-3} \cline{5-7}
		&  Topic  &  Lang/Nation  &   &  Corr.  & p-value &  Quality  \\ \hline
		
		\href{https://nuzzel.com/rank}{NuzzelRank}  &  All  &  All  &  97/99  &  \textcolor{blue}{0.55}  &  \textcolor{blue}{3.7E-09}  &  \textcolor{blue}{0.87}  \\
		\href{https://www.onlinecollegecourses.com/2012/12/17/the-best-english-newspapers/}{OnlineCollegeCourse}  &  General  &  English  &  10/10  &  \textcolor{blue}{0.28}  &  \textcolor{r}{4.3E-01}  &  \textcolor{blue}{0.74}  \\\hline
		\href{https://www.forbes.com/sites/berlinschoolofcreativeleadership/2017/02/01/10-journalism-brands-where-you-will-find-real-facts-rather-than-alternative-facts}{Forbes}  &  General  &  U.S.  &  12/12  &  \textcolor{blue}{0.56}  &  \textcolor{r}{5.6E-02}  &  \textcolor{blue}{0.75}  \\
		\href{http://journalism.wikia.com/wiki/Newspaper_quality_rankings}{JournaWiki}  &  General  &  U.S.  &  41/42  &  \textcolor{blue}{0.68}  &  \textcolor{blue}{1.1E-06}  &  \textcolor{blue}{0.55}  \\
		\href{https://www.ranker.com/list/best-u-s-newspapers/ranker-books}{Ranker}  &  General  &  U.S.  &  49/49  &  \textcolor{blue}{0.40}  &  \textcolor{blue}{4.2E-03}  &  \textcolor{blue}{0.55}  \\
		\href{https://blog.feedspot.com/usa_news_websites/}{FeedSpot U.S.}  &  General  &  U.S.  &  97/104  &  \textcolor{blue}{0.95}  &  \textcolor{blue}{6.1E-49}  &  \textcolor{blue}{0.66}  \\
		\href{https://www.allyoucanread.com/american-newspapers/}{AllYouCanRead U.S.}  &  General  &  U.S.  &  28/30  &  \textcolor{blue}{0.60}  &  \textcolor{blue}{7.9E-04}  &  \textcolor{blue}{0.87}  \\\hline
		\href{https://blog.feedspot.com/italian_news_websites/}{FeedSpot Italian}  &  General  &  Italian  &  5/9  &  \textcolor{blue}{0.50}  &  \textcolor{r}{3.9E-01}  &  \textcolor{r}{0.06}  \\
		\href{https://www.allyoucanread.com/italian-newspapers/}{AllYouCanRead Italian}  &  General  &  Italian  &  29/30  &  \textcolor{blue}{0.39}  &  \textcolor{blue}{3.4E-02}  &  \textcolor{r}{0.40}  \\\hline
		\href{https://www.agilitypr.com/resources/top-media-outlets/top-10-canadian-print-outlets/}{Agility PR Solution}  &  General  &  Canadian  &  10/10  &  \textcolor{blue}{0.41}  &  \textcolor{r}{2.4E-01}  &  \textcolor{r}{0.41}  \\
		\href{https://blog.feedspot.com/canadian_news_websites/}{FeedSpot Canadian}  &  General  &  Canadian  &  57/64  &  \textcolor{blue}{0.79}  &  \textcolor{blue}{1.6E-13}  &  \textcolor{blue}{0.72}  \\
		\href{https://www.allyoucanread.com/canada-newspapers/}{AllYouCanRead Canadian}  &  General  &  Canadian  &  30/30  &  \textcolor{blue}{0.72}  &  \textcolor{blue}{8.3E-06}  &  \textcolor{blue}{0.77}  \\\hline
		\href{http://www.bloghug.com/france-news/}{BlogHub}  &  General  &  French  &  20/20  &  \textcolor{blue}{0.29}  &  \textcolor{r}{2.1E-01}  &  \textcolor{blue}{0.72}  \\
		\href{https://blog.feedspot.com/french_news_websites/}{FeedSpot French}  &  General  &  French  &  8/9  &  \textcolor{blue}{0.55}  &  \textcolor{r}{1.6E-01}  &  \textcolor{r}{0.11}  \\
		\href{https://www.allyoucanread.com/french-newspapers/}{AllYouCanRead French}  &  General  &  French  &  29/30  &  \textcolor{blue}{0.64}  &  \textcolor{blue}{1.6E-04}  &  \textcolor{blue}{0.73}  \\\hline
		\href{https://www.deutschland.de/en/topic/knowledge/news}{DeutschLand}  &  General  &  German  &  4/6  &  \textcolor{blue}{1.00}  &  \textcolor{blue}{0.0E+00}  &  \textcolor{r}{0.32}  \\
		\href{https://blog.feedspot.com/german_news_websites/}{FeedSpot German}  &  General  &  German  &  27/30  &  \textcolor{blue}{0.75}  &  \textcolor{blue}{6.2E-06}  &  \textcolor{blue}{0.61}  \\
		\href{https://www.allyoucanread.com/german-newspapers/}{AllYouCanRead German}  &  General  &  German  &  12/14  &  \textcolor{blue}{0.21}  &  \textcolor{r}{5.1E-01}  &  \textcolor{blue}{0.57}  \\\hline
		\href{https://blog.feedspot.com/spanish_news_websites/}{FeedSpot Spanish}  &  General  &  Spanish  &  5/17  &  \textcolor{blue}{0.70}  &  \textcolor{r}{1.9E-01}  &  \textcolor{r}{0.44}  \\
		\href{https://www.allyoucanread.com/spanish-newspapers/}{AllYouCanRead Spanish}  &  General  &  Spanish  &  19/30  &  \textcolor{blue}{0.78}  &  \textcolor{blue}{9.6E-05}  &  \textcolor{blue}{0.53}  \\\hline
		\href{https://www.fluentu.com/blog/russian/learn-russian-news/}{FluentU}  &  General  &  Russian  &  3/7  &  \textcolor{r}{-0.50}  &  \textcolor{r}{6.7E-01}  &  \textcolor{blue}{0.70}  \\
		\href{https://blog.feedspot.com/russian_news_websites/}{FeedSpot Russian}  &  General  &  Russian  &  6/9  &  \textcolor{blue}{0.94}  &  \textcolor{blue}{4.8E-03}  &  \textcolor{r}{0.18}  \\
		\href{https://www.allyoucanread.com/russian-newspapers/}{AllYouCanRead Russian}  &  General  &  Russian  &  27/30  &  \textcolor{blue}{0.52}  &  \textcolor{blue}{5.5E-03}  &  \textcolor{r}{0.48}  \\\hline
		\href{https://www.penceo.com/blog/top-15-most-popular-sports-websites}{Penceo Sport}  &  Sport  &  All  &  12/15  &  \textcolor{blue}{0.84}  &  \textcolor{blue}{6.4E-04}  &  \textcolor{blue}{0.64}  \\
		\href{https://blog.feedspot.com/sports_news_websites/}{FeedSpot Sport}  &  Sport  &  All  &  30/52  &  \textcolor{blue}{0.65}  &  \textcolor{blue}{8.8E-05}  &  \textcolor{r}{0.45}  \\
		\href{https://www.allyoucanread.com/sports/}{AllYouCanRead Sport}  &  Sport  &  All  &  20/24  &  \textcolor{blue}{0.66}  &  \textcolor{blue}{1.6E-03}  &  \textcolor{blue}{0.80}  \\\hline
		\href{https://www.makeuseof.com/tag/10-entertainment-websites-daily-celebrity-news-fix/}{MakeUseOf}  &  Entertain  &  All  &  8/10  &  \textcolor{blue}{0.36}  &  \textcolor{r}{3.9E-01}  &  \textcolor{r}{0.47}  \\
		\href{https://blog.feedspot.com/hollywood_blogs/}{FeedSpot Entertain}  &  Entertain  &  All  &  13/22  &  \textcolor{blue}{0.18}  &  \textcolor{r}{5.7E-01}  &  \textcolor{r}{0.16}  \\
		\href{https://www.allyoucanread.com/entertainment/}{AllYouCanRead Entertain}  &  Entertain  &  All  &  20/24  &  \textcolor{blue}{0.51}  &  \textcolor{blue}{2.1E-02}  &  \textcolor{blue}{0.88}  \\\hline
		\href{http://www.ebizmba.com/articles/business-websites}{eBizMBA}  &  Business  &  All  &  11/15  &  \textcolor{blue}{0.75}  &  \textcolor{blue}{8.5E-03}  &  \textcolor{blue}{0.78}  \\
		\href{https://blog.feedspot.com/business_news_websites/}{FeedSpot Business}  &  Business  &  All  &  39/46  &  \textcolor{blue}{0.88}  &  \textcolor{blue}{8.3E-14}  &  \textcolor{blue}{0.52}  \\
		\href{https://www.allyoucanread.com/business/}{AllYouCanRead Business}  &  Business  &  All  &  25/26  &  \textcolor{blue}{0.49}  &  \textcolor{blue}{1.4E-02}  &  \textcolor{blue}{0.87}  \\\hline
		\href{http://webtoptenz.com/top-10-tech-news-websites/}{WebTopTen}  &  Tech  &  All  &  10/10  &  \textcolor{blue}{0.68}  &  \textcolor{blue}{2.9E-02}  &  \textcolor{blue}{0.76}  \\
		\href{https://blog.feedspot.com/tech_news_websites/}{FeedSpot Tech}  &  Tech  &  All  &  69/84  &  \textcolor{blue}{0.91}  &  \textcolor{blue}{1.4E-26}  &  \textcolor{blue}{0.55}  \\
		\href{https://www.allyoucanread.com/technology/}{AllYouCanRead Tech}  &  Tech  &  All  &  32/32  &  \textcolor{blue}{0.37}  &  \textcolor{blue}{3.6E-02}  &  \textcolor{blue}{0.74}  \\
		
		\hline 
	\end{tabular}
	\caption{Comparisons of \MediaRank\ to 35 expert news rankings.  ``Quality'' measures the normalized \MediaRank\ scores of common sources, with range [0, 1]. 24 rankings are above 0.05-significance level. Their average Spearman coefficient is 0.69, and average ranking quality score is 0.63.}
	\label{tab:comparepublicranking}
	\vspace{-.15in}
\end{table}

	As shown in Table \ref{tab:comparepublicranking}, Of the 1051 distinct sources mentioned in these rankings, 914 (87\%) are tracked in \MediaRank. Fully 34/35 experts exhibit a positive correlation with our rankings. The average Spearman coefficient is 0.57, and average ranking quality score is 0.58. For rankings with p-value < 0.05 (24 rankings marked blue), the average Spearman coefficient is 0.69, and average ranking quality score is 0.63.
	
	\begin{table*}[!t]
		\footnotesize
		\centering
		\setlength{\tabcolsep}{0.3em}
		\begin{tabular}{@{}|rr|rr|rr|rr|rr|r@{}} \cline{1-10}
			General && Sport && Business && Entertainment && Technology && \multirow{11}{*}{\centering \includegraphics[width=0.085\textwidth]{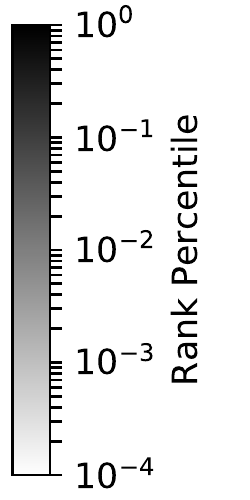}}\\ \cline{1-10}
			
			\textbf{nytimes.com}&\includegraphics[width=0.045\textwidth]{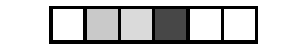} & \textbf{espn.com}&\includegraphics[width=0.045\textwidth]{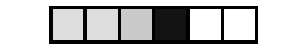} & \textbf{bloomberg.com}&\includegraphics[width=0.045\textwidth]{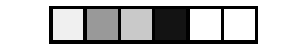} & hollywoodreporter.com&\includegraphics[width=0.045\textwidth]{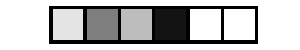} & \textbf{theverge.com}&\includegraphics[width=0.045\textwidth]{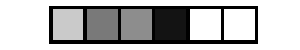} & \\
			\textbf{washingtonpost.com}&\includegraphics[width=0.045\textwidth]{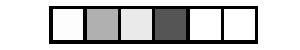} & sbnation.com&\includegraphics[width=0.045\textwidth]{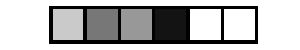} & \textbf{wsj.com}&\includegraphics[width=0.045\textwidth]{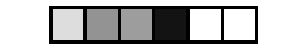} & \textbf{variety.com}&\includegraphics[width=0.045\textwidth]{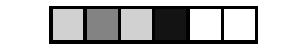} & \textbf{techcrunch.com}&\includegraphics[width=0.045\textwidth]{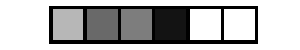} & \\
			\textbf{theguardian.com}&\includegraphics[width=0.045\textwidth]{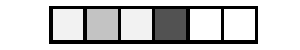} & mlb.com&\includegraphics[width=0.045\textwidth]{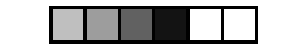} & \textbf{businessinsider.com}&\includegraphics[width=0.045\textwidth]{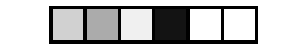} & people.com&\includegraphics[width=0.045\textwidth]{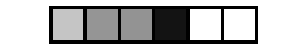} & \textbf{wired.com}&\includegraphics[width=0.045\textwidth]{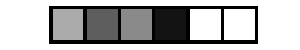} & \\
			\textbf{cnn.com}&\includegraphics[width=0.045\textwidth]{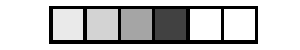} & nfl.com&\includegraphics[width=0.045\textwidth]{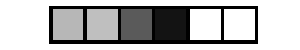} & \textbf{forbes.com}&\includegraphics[width=0.045\textwidth]{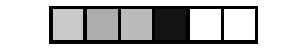} & deadline.com&\includegraphics[width=0.045\textwidth]{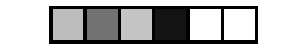} & \textbf{recode.net}&\includegraphics[width=0.045\textwidth]{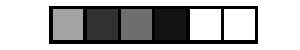} & \\
			\textbf{bbc.com}&\includegraphics[width=0.045\textwidth]{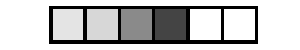} & fifa.com&\includegraphics[width=0.045\textwidth]{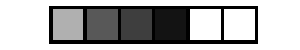} & \textbf{cnbc.com}&\includegraphics[width=0.045\textwidth]{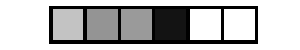} & tmz.com&\includegraphics[width=0.045\textwidth]{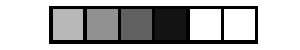} & \textbf{cnet.com}&\includegraphics[width=0.045\textwidth]{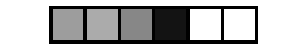} & \\
			\textbf{reuters.com}&\includegraphics[width=0.045\textwidth]{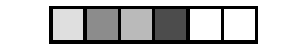} & si.com&\includegraphics[width=0.045\textwidth]{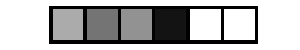} & \textbf{ft.com}&\includegraphics[width=0.045\textwidth]{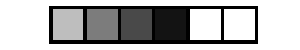} & ew.com&\includegraphics[width=0.045\textwidth]{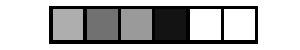} & \textbf{arstechnica.com}&\includegraphics[width=0.045\textwidth]{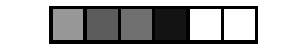} & \\
			\textbf{usatoday.com}&\includegraphics[width=0.045\textwidth]{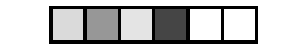} & skysports.com&\includegraphics[width=0.045\textwidth]{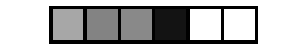} & \textbf{hbr.org}&\includegraphics[width=0.045\textwidth]{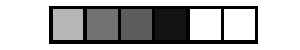} & billboard.com&\includegraphics[width=0.045\textwidth]{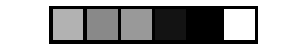} & \textbf{engadget.com}&\includegraphics[width=0.045\textwidth]{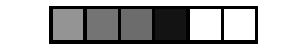} & \\
			\textbf{politico.com}&\includegraphics[width=0.045\textwidth]{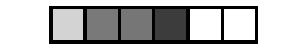} & baseball-reference.com&\includegraphics[width=0.045\textwidth]{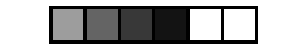} & \textbf{marketwatch.com}&\includegraphics[width=0.045\textwidth]{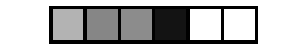} & \textbf{rollingstone.com}&\includegraphics[width=0.045\textwidth]{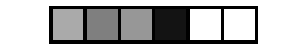} & \textbf{zdnet.com}&\includegraphics[width=0.045\textwidth]{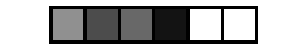} & \\
			\textbf{npr.org}&\includegraphics[width=0.045\textwidth]{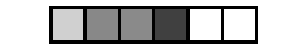} & cbssports.com&\includegraphics[width=0.045\textwidth]{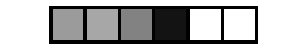} & fool.com&\includegraphics[width=0.045\textwidth]{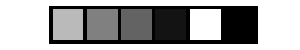} & \textbf{vanityfair.com}&\includegraphics[width=0.045\textwidth]{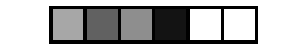} & autonews.com&\includegraphics[width=0.045\textwidth]{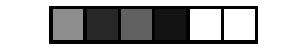} & \\
			\textbf{dailymail.co.uk}&\includegraphics[width=0.045\textwidth]{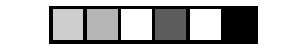} & rotoworld.com&\includegraphics[width=0.045\textwidth]{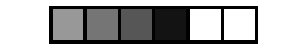} & investopedia.com&\includegraphics[width=0.045\textwidth]{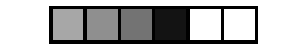} & pagesix.com&\includegraphics[width=0.045\textwidth]{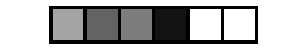} & autocar.co.uk&\includegraphics[width=0.045\textwidth]{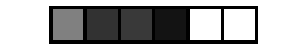} & \\
			
			\cline{1-10} 
		\end{tabular}
		\caption{\MediaRank\ top 10 news of different topics. Sources ranked top in \NuzzelRank\ are shown in bold. Strong agreement in ``General'', ``Business'' and ``Technology''. The rank percentiles of six signals are also visualized (from left to right: reputation, popularity, breadth, bias, social bot and ads scores). Lower ranking sources have lower ranking signals, thus marked in darker color. The \textit{Daily Mail} has strong breadth signal, but it is downgraded due to aggressive ads display.}
		\label{tab:top10_topic}
		\vspace{-.25in}
	\end{table*}

	Table \ref{tab:top10_topic} shows presents the top ten news sources by \MediaRank\ in each of five topic domains.
The sources that also appear on \textit{NuzzelRank}'s top 99 list are highlighted in bold.
There is general agreement between the two systems,
particularly among General, Business and Technology. The range of signal ranking percentile is $[0.0001,
1]$. The smaller the percentile value is, the better quality a source has regarding the signal. Bias is assigned zero for non-political news. We can see that the lower ranking news have darker color. The \textit{Daily Mail} has large breadth, reputation and popularity scores, but its ranking is downgraded due to aggressive ads display.

	\section{Conclusions}
	\label{sec:conclusion}
	
We have demonstrated that the quality of news sources can be instructively
measured using a mix of computational signals reflecting the peer reputation,
reporting bias, bottomline pressure, and popularity.
Our immediate focus now revolves around engineering improvements to our
article analysis, such as 
improved non-English language support for political bias measurement,
e.g. Russian, Chinese and Japanese.
We are also working on improved visualization techniques for news analysis,
to be reflected at \url{www.media-rank.com}.

Deeper NLP analysis of articles to verify or dispute factual claims is a
longer-term goal of this work.   The data collected and released over the course of our
\MediaRank\ project will be a valuable asset to such work.

\begin{acks}
	We thank Prof. Michael Ferdman's help for setting up the server cluster. We are grateful for efforts of Charuta Pethe, Ankur Rastogi, Mohit Goel, Harsh Agarwal, Rohit Patil, Abhishek Reddy in building and maintaining the demonstration website. This work was partially supported by NSF grant IIS-1546113. Any conclusions expressed in this material are of the authors' and do not necessarily reflect the views, either expressed or implied, of the funding party.
\end{acks}

	\bibliographystyle{ACM-Reference-Format}
	\bibliography{sample-bibliography}

\end{document}